\documentclass[12pt]{article}
\usepackage{graphicx}
\usepackage{color}

\def\hybrid{\topmargin 0pt      \oddsidemargin 0pt
        \headheight 0pt \headsep 0pt
       \voffset-1cm
        \textwidth 6.25in       
       \textheight 9.5in       
        \marginparwidth 0.0in
        \parskip 5pt plus 1pt   \jot = 1.5ex}
\catcode`\@=11
\def\marginnote#1{}

\newcount\hour
\newcount\minute
\newtoks\amorpm
\hour=\time\divide\hour by60
\minute=\time{\multiply\hour by60 \global\advance\minute by-\hour}
\edef\standardtime{{\ifnum\hour<12 \global\amorpm={am}%
        \else\global\amorpm={pm}\advance\hour by-12 \fi
        \ifnum\hour=0 \hour=12 \fi
        \number\hour:\ifnum\minute<10 0\fi\number\minute\the\amorpm}}
\edef\militarytime{\number\hour:\ifnum\minute<10 0\fi\number\minute}

\def\draftlabel#1{{\@bsphack\if@filesw {\let\thepage\relax
   \xdef\@gtempa{\write\@auxout{\string
      \newlabel{#1}{{\@currentlabel}{\thepage}}}}}\@gtempa
   \if@nobreak \ifvmode\nobreak\fi\fi\fi\@esphack}
        \gdef\@eqnlabel{#1}}
\def\@eqnlabel{}
\def\@vacuum{}
\def\draftmarginnote#1{\marginpar{\raggedright\scriptsize\tt#1}}

\def\draftlabel#1{{\@bsphack\if@filesw {\let\thepage\relax
   \xdef\@gtempa{\write\@auxout{\string
      \newlabel{#1}{{\@currentlabel}{\thepage}}}}}\@gtempa
   \if@nobreak \ifvmode\nobreak\fi\fi\fi\@esphack}
        \gdef\@eqnlabel{#1}}
\def\@eqnlabel{}
\def\@vacuum{}
\def\draftmarginnote#1{\marginpar{\raggedright\scriptsize\tt#1}}

\def\draft{\oddsidemargin -.5truein
        \def\@oddfoot{\sl preliminary draft \hfil
        \rm\thepage\hfil\sl\today\quad\militarytime}
        \let\@evenfoot\@oddfoot \overfullrule 3pt
        \let\label=\draftlabel
        \let\marginnote=\draftmarginnote
   \def\@eqnnum{(\theequation)\rlap{\kern\marginparsep\tt\@eqnlabel}%
\global\let\@eqnlabel\@vacuum}  }

\def\numberbysection{\@addtoreset{equation}{section}
        \def\theequation{\thesection.\arabic{equation}}}

\def\underline#1{\relax\ifmmode\@@underline#1\else
        $\@@underline{\hbox{#1}}$\relax\fi}

\def\titlepage{\@restonecolfalse\if@twocolumn\@restonecoltrue\onecolumn
     \else \newpage \fi \thispagestyle{empty}\c@page\z@
        \def\thefootnote{\fnsymbol{footnote}} }

\def\endtitlepage{\if@restonecol\twocolumn \else  \fi
        \def\thefootnote{\arabic{footnote}}
        \setcounter{footnote}{0}}  
\relax

\numberbysection
\hybrid

\newfont{\Bbb}{msbm10 scaled 1\@ptsize00}
\newfont{\Bbbb}{msbm7 scaled 1\@ptsize00}

\newcommand{\DDD}{\raise-1pt\hbox{$\mbox{\Bbbb D}$}}

\newcommand{\UUU}{\raise-1pt\hbox{$\mbox{\Bbbb U}$}}

\newcommand{\z}{\raise-1pt\hbox{$\mbox{\Bbbb Z}$}}

\def\beq{\begin{equation}}
\def\eeq{\end{equation}}
\def\p{\partial}

\begin{document}

\begin{titlepage}

\title{Elliptic solutions to integrable nonlinear equations and many-body systems}

\author{A.~Zabrodin\thanks{National Research University Higher School of Economics,
20 Myasnitskaya Ulitsa, Moscow 101000, Russian Federation;
Steklov Mathematical Institute of Russian Academy of Sciences,
Gubkina str. 8, Moscow, 119991, Russian Federation;
e-mail: zabrodin@itep.ru}
}

\date{May 2019}
\maketitle

\vspace{-7cm} \centerline{ \hfill ITEP-TH-08/19}\vspace{7cm}

\begin{abstract}

We review elliptic solutions to integrable nonlinear partial differential 
and difference equations (KP, mKP, BKP, Toda) and derive equations of motion 
for poles of the solutions. The pole dynamics is given by an integrable many-body
system (Calogero-Moser, Ruijsenaars-Schneider). 
The basic tool is the auxiliary linear problems for the wave function
which yield equations of motion together with their Lax representation.
We also discuss integrals of motion and properties of the spectral curves. 

\end{abstract}

\end{titlepage}

\tableofcontents

\vspace{5mm}

\section{Introduction}

The study of singular solutions
to integrable nonlinear partial differential equations and dynamics
of their poles was initiated by the
seminal paper \cite{AMM77}, where elliptic and rational solutions to the
Korteweg-de Vries and Boussinesq equations were investigated. 
It was discovered there that the poles move as particles of the 
integrable many-body Calogero-Moser system \cite{Calogero71,Calogero75,Moser75} 
with some additional restrictions in the phase space.
Later in \cite{Krichever78,CC77} it was shown that in the case of the 
more general Kadomtsev-Petviashvili (KP)
equation the connection with many-body systems becomes most natural: the dynamics of 
poles of rational solutions to the
KP equation is isomorphic to the Calogero-Moser system of particles
with rational pairwise interaction potential
$1/(x_i-x_j)^2$.

Elliptic (double periodic in the complex plane) solutions 
to the KP equation were studied by Krichever in \cite{Krichever80}, where
it was shown that
poles $x_i$ of the elliptic solutions move according to the equations of motion
\beq\label{int1}
\ddot x_i=4\sum_{k\neq i} \wp ' (x_i-x_k)
\eeq
of the Calogero-Moser system of particles with the elliptic
interaction potential $\wp (x_i-x_j)$ ($\wp$ is the Weierstrass $\wp$-function).
Here dot means derivative with respect to the time $t_2$.
The method suggested by Krichever consists in substituting the elliptic solution not in the 
KP equation itself but in the auxiliary linear problems for it, using a suitable
pole ansatz for the wave function depending on a spectral parameter. 
The wave function is defined by its poles and residues at the poles. This method
allows one to obtain the equations of motion together with the Lax representation for them:
\beq\label{int1a}
\dot L=[M, L],
\eeq
where matrices $L$, $M$ depend on $x_i$ and $\dot x_i$ and on the spectral parameter.

The further development is Shiota's work \cite{Shiota94}, where it was shown that the
correspondence between rational solutions to the KP equation and the Calogero-Moser
system with rational potential (when $\wp (x)$ degenerates to $1/x^2$) can be extended
to the level of {\it hierarchies}. There it was proved that the evolution of poles
with respect to the higher times $t_k$ of the infinite KP hierarchy is governed by
higher Hamiltonians $H_k$ of the integrable Calogero-Moser system. 

Another way to derive the equations of motion for poles of singular solutions 
to the KP equation was suggested in
\cite{ABW09}. It consists in parametrizing the 
wave function through its poles $x_i$ and zeros $y_i$, which is basically equivalent to 
substitution of the pole ansatz to the modified KP (mKP) equation. This 
does not allow one to derive the Lax representation but instead 
leads to the so-called self-dual 
form of equations of motion \cite{ABW09}.

Elliptic solutions to the matrix KP equation were studied in \cite{KBBT95}; they 
give rise to the spin generalization of the Calogero-Moser system. Spin degrees
of freedom are related to the matrix residues at the poles of the solutions (which are
fixed in the scalar case). In this paper we will restrict ourselves by the scalar 
(one-component) case.

Solutions of the $B$-version of the KP equation (BKP), which are 
elliptic functions of the time variable $x=t_1$, were recently
discussed in \cite{RZ18}. It was shown that the poles move according to the
following equations of motion:
\beq\label{int8}
\ddot x_i +6\sum_{j\neq i}(\dot x_i +\dot x_j)\wp '(x_i-x_j)-72\!\!
\sum_{j\neq k \neq i}\wp (x_i-x_j)\wp '(x_i-x_k)=0.
\eeq
Here dot means derivative with respect to the time $t_3$.
Instead of the Lax representation, these equations of motion admit a matrix
representation of the form
\beq\label{int8a}
\dot L=[M, L]+P(L-\Lambda I ),
\eeq
where $P$ is a traceless matrix, $I$ is the unity matrix
and $\Lambda$ is the eigenvalue of the Lax matrix $L$
depending on the spectral parameter (Manakov's triple representation \cite{Manakov}).
Matrix elements of the 
matrices $L$, $M$ depend on $x_i$, $\dot x_i$ and on the spectral parameter.

Poles $x_i$ of solutions to the 2D Toda lattice (2DTL), which are elliptic functions of the
discrete time variable $x=t_0$, satisfy the elliptic Ruijsenaars-Schneider
many-body 
system \cite{RS86} (a relativistic version of the Calogero-Moser system):
\beq\label{int2}
\ddot x_i=\sum_{j\neq i}\dot x_i \dot x_j \frac{\wp '(x_i\! -\! x_j)}{\wp (\eta )-
\wp (x_i\! -\! x_j)}
\eeq
(see \cite{KZ95}). Here dot means derivative with respect to the time $t_1$.
In the limit $\eta \to 0$ one recovers the Calogero-Moser system.
The Ruijsenaars-Schneider system is known to be integrable. Similarly to the Calogero-Moser
model, it admits a Lax representation which can be obtained by substituting the 
pole ansatz for the wave function into the semi-difference auxiliary linear problem.
Another approach to the connection between special solutions to the KP and Toda
equations and integrable many-body systems was developed in \cite{Ruij97}.
A self-dual form of the equations of motion also exists (see \cite{ZZ18}).
It is closely connected with the integrable discrete time version of the 
Ruijsenaars-Schneider system \cite{NRK96}, with equations of motion having the form
of the nested Bethe ansatz equations.

This paper is a review of the results mentioned above and
related topics. We give a detailed derivation
of the equations of motion for poles of elliptic solutions to the KP, mKP, BKP and 2DTL
equations together with their Lax (or Manakov's triple) representation and construction
of the spectral curve. Most of the material is contained in the existing literature in one
or another form. However, some points are new. Among them is 
the derivation of the $t_3$-dynamics of poles
of elliptic solutions to the KP equation and equations of motion for poles of 
elliptic solutions to the Novikov-Veselov equation (which is a member of the
2-component BKP hierarchy).

\section{Elliptic solutions to the KP equation}

\subsection{The KP equation}

The KP equation is the first member of an infinite KP hierarchy of 
partial differential equations with independent variables (``times'') 
$t_1=x, t_2, t_3, t_4, \ldots$ \cite{DJKM83,JM83}. The KP equation is
equivalent to the Zakharov-Shabat (``zero curvature'') condition
$\p_{t_3}A_2 -\p_{t_2}A_3 +[A_2, A_3]=0$ for the differential operators
\beq\label{kp1}
A_2=\p_x^2 +2u, \qquad A_3=\p_x^3 +3u\p_x +w.
\eeq
It has the form of a system for two dependent variables $u, w$:
\beq\label{kp2}
\left \{ \begin{array}{l}
2w_x=3u_{t_2}+3u_{xx}
\\ \\
2u_{t_3}-w_{t_2}=6uu_x +2u_{xxx}-w_{xx}.
\end{array}
\right.
\eeq
Excluding $w$ from this system, one obtains the KP equation for the function $u$:
\beq\label{kp3}
3u_{t_2 t_2}=\Bigl (4u_{t_3}-12uu_x -u_{xxx}\Bigr )_x.
\eeq
The Zakharov-Shabat equation (and, therefore, the KP equation) is the compatibility
condition for the auxiliary linear problems
\beq\label{kp3a}
\p_{t_2}\psi =A_2\psi , \qquad \p_{t_3}\psi =A_3\psi
\eeq
for the wave function $\psi$ which depends on a spectral parameter $z$.

The general solution to the whole KP hierarchy is given in terms of the
tau-function $\tau = \tau (t_1, t_2, t_3, \ldots )$.
The change of the dependent variables from $u,w$ to the tau-function
\beq\label{kp4}
u=\p_x^2\log \tau , \qquad w=\frac{3}{2}(\p_x^3\log \tau +\p_{t_2}\p_x \log \tau )
\eeq
brings equations (\ref{kp2}) into the bilinear form \cite{JM83}
\beq\label{kp5}
\Bigl ( D_1^4 -4D_1D_3 +3D_2^2 \Bigr )\tau \cdot \tau =0,
\eeq
where $D_i$ are the Hirota operators. Their action is defined by the rule
$$
P(D_1, D_2, D_3, \ldots )f(t_i) \cdot g(t_i) =
P(\p_{y_1}, \p_{y_2}, \p_{y_3}, \ldots )
f (t_i+y_i)g (t_i-y_i)\Bigr |_{y_i=0}
$$
for any polynomial $P(D_1, D_2, D_3, \ldots )$. The solution 
to the linear equations (\ref{kp3a}) can be expressed through the tau-function
according to the formula \cite{DJKM83}
\beq\label{kp6}
\psi = A(z)\exp \, \Bigl (\, \sum_{k\geq 1}t_k z^k \Bigr )\,
\frac{\tau \Bigr (t_1 -z^{-1}, t_2-\frac{1}{2}\, z^{-2}, t_3-\frac{1}{3}\, z^{-3},
\ldots \Bigr )}{\tau (t_1, t_2, t_3, \ldots )}.
\eeq
Here $z$ is the spectral parameter and $A(z)$ is a normalization factor.

\subsection{The $t_2$-dynamics of poles of elliptic solutions}

Our aim is to study double-periodic (elliptic) in the variable $x$ 
solutions of the KP equation. For such solutions the tau-function is an 
``elliptic quasi-polynomial'' in the variable $x$:
\beq\label{ell1}
\tau = e^{Q(x, t_2, t_3, \ldots )}\prod_{i=1}^{N}\sigma (x-x_i)
\eeq
where $Q(x, t_2, t_3, \ldots )$ is a quadratic form in the times $t_i$ and
$$
\sigma (x)=\sigma (x |\, \omega , \omega ')=
x\prod_{s\neq 0}\Bigl (1-\frac{x}{s}\Bigr )\, e^{\frac{x}{s}+\frac{x^2}{2s^2}},
\quad s=2\omega m+2\omega ' m' \quad \mbox{with integer $m, m'$},
$$ 
is the Weierstrass 
$\sigma$-function with quasi-periods $2\omega$, $2\omega '$ such that 
${\rm Im} (\omega '/ \omega )>0$. It is connected with the Weierstrass 
$\zeta$- and $\wp$-functions by the formulas $\zeta (x)=\sigma '(x)/\sigma (x)$,
$\wp (x)=-\zeta '(x)=-\p_x^2\log \sigma (x)$.
We set $Q=cx^2+bt_2 x +\ldots$ with some constants $b,c$.
The roots $x_i$ are assumed to be 
all distinct. Correspondingly, 
the function $u=\p_x^2\log \tau$ is an 
elliptic function with double poles at the points $x_i$:
\beq\label{ell2}
u=-\sum_{i=1}^{N}\wp (x-x_i) +2c.
\eeq
The poles depend on the times $t_2$, $t_3$.

According to Krichever's method \cite{Krichever80}, the basic tool for 
studying $t_2$-dynamics of poles is the auxiliary linear problem 
$\p_{t_2}\psi =A_2\psi$ for the 
function $\psi$, i.e.,
\beq\label{ell3}
\p_{t_2} \psi =\p_x^2\psi +2u \psi .
\eeq
Since the coefficient function $u$ is double-periodic, 
one can find double-Bloch solutions $\psi (x)$, i.e., solutions such that 
$\psi (x+2\omega )=b \psi (x)$, $\psi (x+2\omega ' )=b' \psi (x)$
with some Bloch multipliers $b, b'$. Equations (\ref{kp6}), (\ref{ell1}) tell us that the
wave function $\psi$ has simple poles at the points $x_i$.
The pole ansatz for the $\psi$-function is
\beq\label{ell4}
\psi = e^{xz+t_2z^2+t_3z^3}\sum_{i=1}^N c_i \Phi (x-x_i, \lambda ),
\eeq
where the coefficients $c_i$ do not depend on $x$.
Here we use the function
\beq\label{Phi}
\Phi (x, \lambda )=\frac{\sigma (x+\lambda )}{\sigma (\lambda )\sigma (x)}\,
e^{-\zeta (\lambda )x}
\eeq
which has a simple pole
at $x=0$ ($\zeta$ is the Weierstrass $\zeta$-function). 
The expansion of $\Phi$ as $x\to 0$ is
$$
\Phi (x, \lambda )=\frac{1}{x}+\alpha_1 x +\alpha_2 x^2 +\ldots , \qquad 
x\to 0,
$$
where $\alpha_1=-\frac{1}{2}\, \wp (\lambda )$, $\alpha_2=-\frac{1}{6}\, \wp '(\lambda )$. 
The parameters $z$ and $\lambda$ 
are spectral parameters. They are going to be connected by equation of the spectral curve
(see below).
Using the quasiperiodicity properties of the function $\Phi$,
$$
\Phi (x+2\omega , \lambda )=e^{2(\zeta (\omega )\lambda - \zeta (\lambda )\omega )}
\Phi (x, \lambda ),
$$
$$
\Phi (x+2\omega ' , \lambda )=e^{2(\zeta (\omega ' )\lambda - \zeta (\lambda )\omega ' )}
\Phi (x, \lambda ),
$$
one concludes that $\psi$ given by (\ref{ell4}) 
is indeed a double-Bloch function with Bloch multipliers
$$b=e^{2(\omega z + \zeta (\omega )\lambda - \zeta (\lambda )\omega )}, \qquad
b '=e^{2(\omega ' z + \zeta (\omega ' )\lambda - \zeta (\lambda )\omega ' )}.$$
We will often suppress the second argument of $\Phi$ writing simply 
$\Phi (x)=\Phi (x, \lambda )$. 
We will also need the $x$-derivatives  
$\Phi '(x, \lambda )=\p_x \Phi (x, \lambda )$, 
$\Phi ''(x, \lambda )=\p^2_x \Phi (x, \lambda )$,
etc. 

The function $-\p_{t_2}\psi +\p_x^2 \psi +2u\psi$ is also a double-Bloch function
with the same Bloch multipliers. If we manage to prove that this function is free of poles,
then the only possibility for it is the exponential function $Ce^{ax}$, which, however,
has a pair of Bloch multipliers that is not equivalent to $b, b'$. 
Therefore, $C=0$ and the function 
vanishes identically. 
Substituting (\ref{ell4}) into (\ref{ell3}) with $u$ given by (\ref{ell2}), we get:
$$
-\sum_i \dot c_i \Phi (x-x_i)+\sum_i c_i \dot x_i \Phi '(x-x_i)+2z
\sum_i c_i \Phi '(x-x_i)+\sum_i c_i \Phi ''(x-x_i)
$$
$$
-\, 2\left (\sum_i \wp (x-x_i)\right ) \left ( \sum_k c_k \Phi (x-x_k)\right )
+4c \sum_i c_i \Phi (x-x_i)=0,
$$
where dot means the $t_2$-derivative. Different terms of this expression have poles at $x=x_i$. 
The highest poles are of third order but it is easy to see that they cancel identically.
It is a matter of direct calculation to see that the conditions of cancellation of
second and first order poles have the form
\beq\label{ell5}
c_i\dot x_i=-2zc_i -2\sum_{j\neq i}c_j \Phi (x_i-x_j),
\eeq
\beq\label{ell6}
\dot c_i=(4c-2\alpha_1) c_i -2\sum_{j\neq i}c_j \Phi '(x_i-x_j)-2c_i \sum_{j\neq i}
\wp (x_i-x_j).
\eeq
They have to be valid for all $i=1, \ldots , N$. Introducing $N\! \times \! N$ matrices
\beq\label{ell7}
L_{ij}=-\delta_{ij}\dot x_i -2(1-\delta_{ij})\Phi (x_i-x_j),
\eeq
\beq\label{ell8}
M_{ij}=\delta_{ij}(\wp (\lambda )+4c)-2\delta_{ij}
\sum_{k\neq i}\wp (x_i-x_k) -2(1-\delta_{ij})\Phi ' (x_i-x_j),
\eeq
we can write the above conditions as a system of linear equations for the vector
${\bf c}=(c_1, \ldots , c_N)^T$:
\beq\label{ell9}
\left \{ \begin{array}{l}
L{\bf c}=2z{\bf c}
\\ \\
\dot {\bf c}=M {\bf c}.
\end{array} \right.
\eeq
It is convenient to introduce diagonal matrices $I$, $X$, $D$ given by
$I_{ik}=\delta_{ik}$, $X_{ik}=\delta_{ik}x_i$, 
\beq\label{ell11}
D_{ik}=\delta_{ik}\sum_{j\neq i}\wp (x_i-x_j)
\eeq
and off-diagonal matrices
$A$, $B$ given by
\beq\label{ell10}
A_{ik}=(1-\delta_{ik})\Phi (x_i-x_k),
\qquad
B_{ik}=(1-\delta_{ik})\Phi ' (x_i-x_k),
\eeq
then the matrices $L,M$ are
$$
L=-\dot X -2A, \qquad M=(\wp (\lambda )+4c)I-2B-2D.
$$
Differentiating the first equation in (\ref{ell9}) with respect to $t_2$, we arrive at
the compatibility condition of the linear problems (\ref{ell9}):
\beq\label{ell12}
\Bigl (\dot L+[L,M]\Bigr ) {\bf c} =0.
\eeq
We have
$$
\dot L+[L,M]=-\ddot X -2\dot A +2[\dot X, B] +4[A,B]+4[A,D].
$$
It is straightforward to see that $\dot A=[\dot X, B]$. Next, in the appendix it 
is proved that $[A,B]+[A,D]=D'$, where $D'$ is the diagonal matrix
$\displaystyle{D'_{ik}=\delta_{ik}\sum_{j\neq i}\wp '(x_i-x_j)}$. Therefore,
we have the identity
$$
\dot L+[L,M]=-\ddot X +4D',
$$
and so the compatibility condition states that $(-\ddot X +4D')_{ii}=0$ for all 
$i=1, \ldots , N$. This implies the Calogero-Moser 
equations of motion (\ref{int1}) together with their 
Lax representation (\ref{int1a}).

It follows from the Lax representation that the time evolution is an isospectral
transformation of the Lax matrix, so all traces $\mbox{tr}\, L^k$ and the 
characteristic polynomial $\det (L-2zI)$ are integrals of motion.

The Calogero-Moser system is Hamiltonian with the Hamiltonian
\beq\label{ell13}
H_2=\sum_i p_i^2 -2\sum_{i<j}\wp (x_i-x_j)
\eeq
and the Poisson brackets $\{x_i, p_k\}=\delta_{ik}$. Since $\dot x_i=
\p H_2/\p p_i =2p_i$, the Lax matrix expressed through the momenta reads
$L_{ij}=-2\Bigl (\delta_{ij}p_i +(1-\delta_{ij})\Phi (x_i-x_j)\Bigr )$. The Hamiltonian
is given by
\beq\label{ell14}
H_2=\frac{1}{4}\, \mbox{tr}\, L^2 -N(N-1)\wp (\lambda ).
\eeq
The higher Hamiltonians can be defined by the formulas
\beq\label{ell15}
H_k=2^{-k}\mbox{tr}\, L^k +\sum_{m=0}^{k-2}a_m(\lambda )\mbox{tr}\, L^m,
\eeq
where $a_m(\lambda )$ are some elliptic functions of $\lambda $ which are determined
by the requirement that the Hamiltonians should be $\lambda$-independent. In particular,
\beq\label{ell16}
\begin{array}{lll}
H_1&=&\displaystyle{\frac{1}{2}\, \mbox{tr}\, L=-\sum_ip_i ,}
\\ &&\\
H_3&=& \displaystyle{
\frac{1}{8}\, \mbox{tr}\, L^3 -\frac{3}{2}\, (N-1)\wp (\lambda )\mbox{tr}\, L-
\frac{1}{2}\, N(N-1)(N-2)\wp ' (\lambda )}
\\ && \\
&=&\displaystyle{-\sum_i p_i^3 +3\sum_{i\neq j}p_i \, \wp (x_i-x_j)}.
\end{array}
\eeq

\subsection{The $t_3$-dynamics of poles of elliptic solutions}

Similarly to the $t_2$-case, we should substitute $\psi$ given by (\ref{ell4}),
into the linear problem
$$
\p_{t_3}\psi =\p_x^3\psi +3u\p_x \psi +w\psi .
$$
According to (\ref{kp4}), (\ref{ell1}), we have for elliptic solutions:
\beq\label{t31}
w=-\frac{3}{2}\sum_i \wp '(x-x_i)+\frac{3}{2}\sum_i\dot x_i \wp (x-x_i)+\frac{3}{2}\, b.
\eeq
The substitution leads to the following equation:
$$
\sum_i\p_{t_3}c_i\Phi (x-x_i)-\sum_i c_i \p_{t_3}x_i \Phi '(x-x_i)
$$
$$
=3z^2 \sum_i c_i \Phi '(x-x_i)+3z \sum_i c_i \Phi ''(x-x_i)+
\sum_i c_i \Phi '''(x-x_i)
$$
$$
-3z\left (\sum_i \wp (x-x_i)\right )\left (\sum_jc_j \Phi (x-x_j)\right )
-3\left (\sum_i \wp (x-x_i)\right )\left (\sum_jc_j \Phi '(x-x_j)\right )
$$
$$
-\frac{3}{2}\left (\sum_i \wp '(x-x_i)\right )\left (\sum_jc_j \Phi (x-x_j)\right )
+\frac{3}{2}\left (\sum_i \dot x_i
\wp (x-x_i)\right )\left (\sum_jc_j \Phi (x-x_j)\right )
$$
$$
+6cz\sum_i c_i \Phi (x-x_i)+6c \sum_i c_i \Phi '(x-x_i)+\frac{3}{2}\, b
\sum_i c_i \Phi (x-x_i).
$$
The fourth order poles at $x=x_i$ cancel identically. The cancellation of the third
order poles leads to the same condition (\ref{ell5}) which is the eigenvalue 
equation $L{\bf c}=2z{\bf c}$ for the Lax matrix. The cancellation of the second order
poles leads to the equations
\beq\label{t32}
\p_{t_3}x_i c_i=-3z^2c_i -6cc_i -3z\sum_{j\neq i}c_j \Phi (x_i-x_j)
+3c_i \sum_{j\neq i}\wp (x_i-x_j)+\frac{3}{2}\, \dot x_i \sum_{j\neq i}c_j \Phi (x_i-x_j).
\eeq
Taking into account (\ref{ell5}), one can rewrite these equations as
\beq\label{t33}
\p_{t_3}x_i =-6c -\frac{3}{4}\, \dot x_i^2 +3\sum_{j\neq i}\wp (x_i-x_j),
\eeq
or, in the matrix form,
\beq\label{t34}
\p_{t_3}X=-6cI -\frac{3}{4}\, \dot X^2 +3D.
\eeq
Finally, the cancellation of the simple poles at $x=x_i$, with the 
equation $\dot {\bf c}=M{\bf c}$ being taking into account, leads to the conditions
$$
\p_{t_3}c_i=\frac{3}{2}\, z\dot c_i -\frac{3}{2}\sum_{j\neq i}c_j \Phi ''(x_i-x_j)
+\frac{3}{2}\, c_i \sum_{j\neq i}\wp '(x_i-x_j)+\frac{1}{2}\, \wp '(\lambda )c_i +
\frac{3}{2}\, bc_i
$$
$$
+\frac{3}{2}\dot x_i \sum_{j\neq i}c_j \Phi '(x_i-x_j)+\frac{3}{2}\, c_i
\sum_{j\neq i}\dot x_j \wp (x_i-x_j)-\frac{3}{4}\, \wp (\lambda )\dot x_i c_i.
$$
Introducing the diagonal matrix 
$\displaystyle{\tilde D_{ik}=\delta_{ik}\sum_{j\neq i}\dot x_j \wp (x_i-x_j)}$
and the off-diagonal matrix $C_{ik}=(1-\delta_{ik})\Phi '' (x_i-x_k)$, we can rewrite
these conditions in the matrix form as
\beq\label{t35}
\p_{t_3}{\bf c}=T{\bf c},
\eeq
where $T$ is the matrix
\beq\label{t36}
T=\frac{3}{4}\, ML -\frac{3}{2}\, C +\frac{3}{2}\, \dot X B+\frac{3}{2}\, D'
+\frac{3}{2}\, \tilde D -\frac{3}{4}\, \wp (\lambda )\dot X +\frac{1}{2}
(\wp '(\lambda )+3b)I.
\eeq

The compatibility condition of the linear system (\ref{t35}) and the 
equation $L{\bf c}=2z{\bf c}$ is $(\p_{t_3}L+[L, T]){\bf c}=0$. Let us calculate
$\p_{t_3}L+[L, T]$ using the Lax equation $\dot L=[M,L]$. We have, after some algebra:
$$
\p_{t_3}L\! +\! [L, T]=-\p_{t_3}\dot X \! -\! 
3D'\dot X +3\Bigl ([A,C]+2[B, D]-AD'\! -\! D'A \Bigr )+3Y
+\frac{3}{2}[\dot X, C-\wp (\lambda )A],
$$
where $Y=B\dot X A-A\dot X B -[A, \tilde D]$. In the appendix we prove the matrix
identity $[A,C]+2[B, D]-AD'\! -\! D'A=0$. Moreover, it holds
$$
Y_{ik}+\frac{1}{2}[\dot X, C-\wp (\lambda )A]_{ik}=0 \qquad \mbox{for $i\neq k$}
$$
and
$$
Y_{ii}=-\sum_{j\neq i}\dot x_j \wp '(x_i-x_j)
$$
(see (\ref{B1}) in the appendix). Therefore, we obtain that the matrix 
$\p_{t_3}L\! +\! [L, T]$ is diagonal and
\beq\label{t37}
\Bigl (\p_{t_3}L\! +\! [L, T]\Bigr )_{ii}=-\p_{t_3}\dot x_i -3
\sum_{j\neq i}(\dot x_i+\dot x_j)\wp '(x_i-x_j).
\eeq
This means that the compatibility condition leads to the equations of motion
\beq\label{t38}
\p_{t_3}\dot x_i +3
\sum_{j\neq i}(\dot x_i+\dot x_j)\wp '(x_i-x_j)=0.
\eeq
As one can easily see, these equations coincide with the $t_2$-derivatives of 
equations (\ref{t33}). Moreover, equations (\ref{t33}) and (\ref{t38}), when rewritten
in terms of the moments $p_i$, are Hamiltonian equations for the $\tilde H_3$
Hamiltonian flow (here $\tilde H_3=H_3+6cH_1$):
\beq\label{t39}
\left \{ \begin{array}{l}
\displaystyle{\p_{t_3} x_i =\frac{\p \tilde H_3}{\p p_i}=-6c -3p_i^2+3
\sum_{j\neq i}\wp (x_i-x_j)}
\\ \\
\displaystyle{\p_{t_3} p_i =-\frac{\p \tilde H_3}{\p x_i}=-3
\sum_{j\neq i}(p_i+p_j)\wp ' (x_i-x_j)}
\end{array}
\right.
\eeq
($H_3$ is given in (\ref{ell16})).

It is easy to see that the result of \cite{AMM77} for the KdV equation
($u$ is independent of $t_2$) follows from (\ref{t39}): the equations of motion are
$\displaystyle{\p_{t_3} x_i =-6c +3
\sum_{j\neq i}\wp (x_i-x_j)}$ on the locus 
defined by the equations $\displaystyle{\sum_{j\neq i}\wp '(x_i-x_j)=0}$.

\subsection{Reviving the coupling constant}

The Calogero-Moser system that is obtained form the dynamics of poles
comes with a fixed coupling constant. A natural question is whether other values
of the coupling constant are possible. It appears that in order to revive
the coupling constant one should consider the so-called $\hbar$-version of the 
KP hierarchy ($\hbar$-KP) instead of the standard KP one. The $\hbar$-KP hierarchy
is obtained by the formal change of times $t_k \to t_k /\hbar$, so that the $\hbar$-KP
equation acquires the form
\beq\label{h1}
3 u_{t_2 t_2}=\Bigl (4u_{t_3}-12uu_x -\hbar^2 u_{xxx}\Bigr )_x
\eeq
and the second linear problem is
\beq\label{h2}
\hbar \p_{t_2}\psi =\hbar^2 \p_x^2\psi +2\hbar^2 \p_x^2 \log \tau \, \psi .
\eeq
Substituting the wave function in the form
\beq\label{h3}
\psi = e^{\frac{1}{\hbar}\, xz+\frac{1}{\hbar}\, t_2 z^2 + \frac{1}{\hbar}\, t_3 z^3}
\sum_i c_i \Phi (x-x_i )
\eeq
with the same expression for 
$\tau$ (\ref{ell1}), we get, after writing down conditions of cancellation of the poles,
the same system (\ref{ell9}) with the matrices
\beq\label{h4}
L_{ij}=-\delta_{ij}\dot x_i -2\hbar (1-\delta_{ij})\Phi (x_i-x_j),
\eeq 
\beq\label{h5}
M_{ij}=(\hbar \wp (\lambda ) +4c\hbar^{-1})\delta_{ij}-2\hbar \delta_{ij}
\sum_{k\neq i}\wp (x_i-x_k)-2\hbar (1-\delta_{ij}) \Phi ' (x_i-x_j).
\eeq
Then, repeating the calculation above, one obtains 
$$
\dot L+[L,M]=-\ddot X +4\hbar^2 D'
$$
and the equations of motion acquire the form
\beq\label{h6}
\ddot x_i =4\hbar^2 \sum_{j\neq i} \wp ' (x_i-x_j)
\eeq
with the coupling constant $4\hbar^2$.

\subsection{The spectral curve}

From now on we put $\hbar =1$, as before. 
The equation of the spectral curve is
\beq\label{spec1}
R(z, \lambda )=\det \Bigl (2zI-L(z, \lambda )\Bigr )=0.
\eeq
As it was already mentioned, the equation of the spectral curve 
(the characteristic equation of the Lax matrix) is an integral of motion
by virtue of the Lax equation for both $t_2$ and $t_3$ flows: 
$\frac{d}{dt_2}\det \Bigl (2zI-L(z, \lambda )\Bigr )=
\frac{d}{dt_3}\det \Bigl (2zI-L(z, \lambda )\Bigr )=0$. 

The matrix
$L=L(z, \lambda )$, which has essential singularity at $\lambda =0$, can be 
represented in the form $L=G\tilde L G^{-1}$, where matrix elements of 
$\tilde L$ do not have 
essential singularities and $G$ is the diagonal matrix $G_{ij}=\delta_{ij}
e^{-\zeta (\lambda )x_i}$. Therefore, 
$$
R(z, \lambda )=\sum_{k=0}^{N}R_k(\lambda )z^k,
$$
where the coefficients $R_k(\lambda )$ are elliptic functions of $\lambda$ with poles
at $\lambda =0$.
The functions 
$R_k (\lambda )$ can be represented as linear combinations of $\wp$-function and
its derivatives. Coefficients
of this expansion are integrals of motion. Fixing values of these integrals, we obtain
via the equation $R(z, \lambda )=0$ the algebraic curve $\Gamma$ which is a 
$N$-sheet covering of the initial elliptic curve ${\cal E}$ realized as a factor
of the complex plane with respect to the lattice generated by $2\omega$, $2\omega '$. 

\noindent
{\it Example} ($N=2$):
$$
\det_{2\times 2}\Bigl (2zI-L(z, \lambda )\Bigr )=4z^2 +2z(\dot x_1 +\dot x_2)
+\dot x_1 \dot x_2 +4\wp (x_1-x_2)-4\wp (\lambda ).
$$
{\it Example} ($N=3$):
$$
\det_{3\times 3}\Bigl (2zI-L(z, \lambda )\Bigr )=8z^3+4z^2 (\dot x_1 +\dot x_2 +\dot x_3)
$$
$$
+2z \Bigl (\dot x_1\dot x_2 +\dot x_1\dot x_3+ \dot x_2\dot x_3+
4\wp (x_{12})+4\wp (x_{13})+4\wp (x_{23})-12\wp (\lambda )\Bigr )
$$
$$
+\dot x_1 \dot x_2 \dot x_3 +4\dot x_1\wp (x_{23})+4\dot x_2\wp (x_{13})+
4\dot x_3\wp (x_{12})-4\wp (\lambda )(\dot x_1 +\dot x_2 +\dot x_3)-8\wp '(\lambda ),
$$
where $x_{ik}=x_i-x_k$. 

In a neighborhood of $\lambda =0$   
the matrix $\tilde L$
can be written as
$$
\tilde L=-2\lambda ^{-1}(E-I)+O(1),
$$ 
where
$E$ is the rank $1$ matrix with matrix elements $E_{ij}=1$ for all $i,j =1, \ldots , N$.
The matrix $E$ has eigenvalue $0$ with multiplicity $N-1$ and another eigenvalue 
equal to $N$. Therefore, we can write $R(z, \lambda )$ in the form
\beq\label{spec2}
\begin{array}{lll}
R(z, \lambda )&=&
\det \Bigl (2z+2\lambda ^{-1}(E-I)+O(1)\Bigr )
\\ && \\
&=&\displaystyle{2^N \Bigl (z+(N\! -\! 1)\lambda ^{-1}-f_N(\lambda )\Bigr )\prod_{i=1}^{N-1}
(z-\lambda ^{-1}-f_i(\lambda ))},
\end{array}
\eeq
where $f_i$ are regular functions of $\lambda$ at $\lambda =0$: $f_i(\lambda )=O(1)$ as
$\lambda \to 0$. 
This means that the function $z$
has simple poles on all sheets at the points $P_j$ ($j=1, \ldots , N$) of the curve
$\Gamma$ located 
above $\lambda =0$. Its expansion in the
local parameter $\lambda$ on the sheets near these points is given by the multipliers
in the right hand side of (\ref{spec2}).  So we have the following expansions 
of the function $z$ near 
the ``points at infinity'' $P_j$:
\beq\label{spec2a}
\begin{array}{l}
z=\, \lambda^{-1}+f_j(\lambda ) \quad \mbox{near $P_j$}, \quad j=1, \ldots , N-1,
\\ \\
z=-(N\! -\! 1)\lambda ^{-1}+f_{N}(\lambda ) \quad \mbox{near $P_{N}$}.
\end{array}
\eeq
The $N$-th sheet is distinguished, as it can be seen 
from (\ref{spec2}), (\ref{spec2a}). As in \cite{Krichever80}, we call it the upper sheet.

The genus $g$ 
of the spectral curve $\Gamma$ can be found using 
the following argument. Let us apply the Riemann-Hurwitz formula
to the covering $\Gamma \to {\cal E}$. We have $2g-2=\nu$, where $\nu$ is the number 
of ramification points of the covering, which are zeros on $\Gamma$
of the function $\p R/\p z$. Differentiating equation (\ref{spec2}) with respect to $z$, 
we can conclude that the function $\p R/\p z$ has simple poles at the points $P_1,
\ldots , P_{N-1}$ on all
sheets except the upper one, where it has a pole of order $N-1$. The number
of poles of any meromorphic function is equal to the number of zeros. Therefore,
$\nu = 2(N-1)$ and so $g=N$.

The spectral curve $\Gamma$ is not smooth because in general position the genus
of the curve which is a $N$-sheet covering of an elliptic curve is
$g=\frac{1}{2}N(N-1)+1$.

\subsection{The $\psi$-function as the Baker-Akhiezer function on the spectral curve}
\label{section:ba}

Let $P$ be a point of the curve $\Gamma$, i.e. $P=(z, \lambda )$, where 
$z$ and $\lambda$ are connected by the equation $R(z, \lambda )=0$. The coefficients
$c_i$ in the pole ansatz for the function $\psi$, after normalization, 
are functions on the curve $\Gamma$:
$c_i=c_i(t_2, P)$. Let us normalize them by the condition $c_1(0, P)=1$.
The non-normalized components $c_i(0, P)$ are equal to 
$\Delta_i (0, P)$, where $\Delta_i (0, P)$ are suitable minors of the matrix
$2zI-L(0)$. They are holomorphic functions on $\Gamma$ outside the
points above $\lambda =0$. After normalizing the first component, all other components
$c_i(0,P)$ become meromorphic functions on $\Gamma$ outside the points $P_j$ located
above $\lambda =0$. Their poles are zeros on $\Gamma$ of the first minor of the matrix
$2zI-L(0)$, i.e., they are given by common solutions of 
equation (\ref{spec1}) and the equation 
$\det \Bigl (2z\delta_{ij}-L_{ij}(0)\Bigr )=0$, 
$i,j=2, \ldots , N$. The location of these poles depends on the initial data.

The number of the poles can be found by the following argument \cite{Krichever80,KBBT95}. 
Let us consider
the function $F$ of the complex variable $\lambda \in {\cal E}$ defined by
$$
F(\lambda )=\Bigl (\det c_i(0, P_j(\lambda ))\Bigr )^2,
$$
where $P_j(\lambda )$ are $N$ pre-images of the point $\lambda$ under the projection
$\Gamma \to {\cal E}$. This function is well-defined as a function of 
$\lambda$ (since it does not depend on the order of the sheets) and
has double poles at the images of the poles of $c_i$'s. Clearly, $F$ only 
vanishes at the ramification points, where at least
two columns of the matrix $c_i(0, P_j(\lambda ))$ coincide. Indeed, let 
$P_j=(z_j, \lambda )$ be the $N$ points above $\lambda$. Then $c_i (P_j)$ are
eigenvectors of $L(\lambda )$ with the eigenvalues $2z_j$. They are linearly independent 
if all the $z_j$'s are different. Therefore, $F$ does not vanish at such a point. 
Let us now assume that $\lambda$ is a ramification point which is generically of order 2.
It is easy to see that at such a point $F$ has a simple zero. Indeed, let $\xi$ be
a local parameter on the curve around the ramification point, then $\lambda =
\lambda_0+\lambda_1 \xi^2 +O(\xi^3)$ in a small neighborhood of this point. The 
determinant is $O(\xi )$ hence $F =O(\xi^2)$ but this is precisely proportional to 
$\lambda -\lambda_0$. If $M$ is the number of poles of the vector ${\bf c}$ on $\Gamma$,
then $2M=\nu =2(N-1)$, hence $M=N-1$.

Finding explicitly eigenvectors of the matrix $E-I$, one can see that
in a neighborhood of the ``points at infinity'' $P_j$ the functions 
$c_i (0, P)$ have the form
\beq\label{spec3}
c_i (0, P)=\Bigl (c_i^{0(j)} +O(\lambda )\Bigr )e^{-\zeta (\lambda )(x_i(0)-x_1(0))},
\quad 2\leq i\leq N, \quad j\neq N,
\eeq
where $\displaystyle{\sum_{i=2}^{N}c_i^{0(j)}=-1}$ and
\beq\label{spec4}
c_i (0, P)=\Bigl (1 +O(\lambda )\Bigr )e^{-\zeta (\lambda )(x_i(0)-x_1(0))},
\quad 2\leq i\leq N, \quad j= N
\eeq
(on the upper sheet).

The fundamental matrix ${\cal S}(t_2)$ of solutions to the equation 
$\p_{t_2} {\cal S} =M{\cal S}$, ${\cal S}(0)=I$, is a regular function of 
$\lambda$ for $\lambda \neq 0$.
From the Lax equation it 
follows that ${\bf c}(t_2)={\cal S}(t_2){\bf c}(0)$ is the 
common solution of the equations $\dot {\bf c}=M{\bf c}$ and $L{\bf c}=2z {\bf c}$.
Thus the vector ${\bf c}(t_2, P)$ has the same $t_2$-independent poles as the vector
${\bf c}(0,P)$.

In order to find $c_i (t_2, P)$ near the pre-images of the point $\lambda =0$ it is convenient
to pass to the gauge equivalent pair $\tilde L$, $\tilde M$, where
$$
\tilde L=G^{-1}LG , \quad \tilde M =-G^{-1}\p_{t_2} G +G^{-1}MG
$$
with the same diagonal matrix $G$ as before. Let $\tilde {\bf c}=G^{-1}{\bf c}$ be the
gauge-transformed vector ${\bf c}=(c_1, \ldots , c_N)^T$, then our linear system is
$$
\tilde L \tilde {\bf c}=2z\tilde {\bf c}, \qquad
\p_t \tilde {\bf c}=\tilde M \tilde {\bf c}.
$$

By a straightforward calculation one can check that the following relation holds:
\beq\label{spec5}
\tilde M=\lambda^{-2}I-\lambda^{-1}\tilde L +O(1).
\eeq
Applying the both sides to the eigenvector $\tilde {\bf c}$ of $\tilde L$
with the eigenvalue $2z$, we get
\beq\label{spec6}
\p_{t_2} \tilde {\bf c}=(\lambda^{-2}-2z\lambda^{-1})\tilde {\bf c}
+O(1).
\eeq
Therefore, since $z=\lambda^{-1}+O(1)$ on all sheets except the upper one, we have
\beq\label{spec7}
\p_{t_2} \tilde {\bf c}^{(j)}=-(z^2+O(1))\tilde {\bf c}^{(j)}, \quad j=1, \ldots , N-1,
\eeq
so 
$$
\tilde {\bf c}^{(j)}(t_2,P)=({\bf c}^{0(j)}+O(\lambda ))
e^{-z^2t_2}, \quad j=1, \ldots , N-1.
$$
For the vector $\tilde {\bf c}^{(N)}$ on the
upper sheet we have from (\ref{spec6}), recalling (\ref{spec2a}), 
\beq\label{spec7a}
\p_{t_2} \tilde {\bf c}^{(N)}=\Bigl 
(-z^2+k^2(\lambda ) +O(1)\Bigr )\tilde {\bf c}^{(N)}, 
\eeq
where
$$
k(\lambda )=-N\lambda^{-1}+f_{N},
$$
so
$$
\tilde {\bf c}^{(N)}(t_2,P)=({\bf e}+O(\lambda ))e^{(-z^2+k^2(\lambda ))t_2} 
$$
(here ${\bf e}=(1, 1, \ldots , 1)^T$).
Coming back to the vector ${\bf c}(t_2,P)$, we obtain after normalization
\beq\label{spec8}
c_i^{(j)}(t_2,P)=c_{ij}(\lambda ) e^{-\zeta (\lambda )(x_i(t_2)-x_1(0))+\nu_j (\lambda )t_2},
\eeq
where $\nu_j =-z^2$ for $j=1, \ldots , N-1$, $\nu_{N}=-z^2 +k^2 (\lambda )$ and
$c_{ij}(\lambda )$ are regular functions in a neighborhood of $\lambda =0$. Their
values at $\lambda =0$ are
\beq\label{spec9}
c_{1j}(0)=1, \quad j=1, \ldots , N, \qquad c_{ij}(0)=c_i^{0(j)}, \quad i\geq 2, \,\,
j\neq N, \qquad c_{i\, N}(0)=1,
\eeq
with $\displaystyle{\sum_{i=2}^{N}c_i^{0(j)}=-1}$.

After investigating the analytic properties of the vector ${\bf c}(t_2,P)$ 
let us turn to the function $\psi$:
$$
\psi (x, t_2, P)=\sum_{i=1}^{N}c_i(t_2, P)\Phi (x-x_i , \lambda )e^{zx+z^2t_2}.
$$
The function $\Phi (x-x_i, \lambda )$ has essential singularities at all points
$P_j$ located above $\lambda =0$. It follows from (\ref{spec8}) that in the function
$\psi$ these essential 
singularities cancel on all sheets except the upper one, where 
$\psi \propto e^{k(\lambda )x+k^2(\lambda )t_2}e^{\zeta (\lambda )x_1(0)}$. 
From (\ref{spec9}) it follows that
$\psi$ has a simple pole at the point $P_{N}$ and no poles at the points 
$P_j$ for $j=1, \ldots N-1$. As we have seen before,
the function $\psi$ also has $N-1$ poles in the
finite part of the curve $\Gamma$, which do not depend on $x,t_2$.
These analytic properties allows one to identify the function 
$\psi (x, t_2, P)$ with the Baker-Akhiezer function on the spectral curve $\Gamma$
with the marked point at infinity $P_N$.

\subsection{Self-dual form of the equations of motion}
\label{section:sdcm}

Another way to derive the equations of motion is to parametrize the wave function
through its poles and zeros and substitute into the linear problem (\ref{ell3}).
Namely, represent the wave function as $\psi =\tilde \tau / \tau$, then 
the linear problem (\ref{ell3}) acquires the form
\beq\label{sd1}
\p_{t_2}\log \frac{\tilde \tau}{\tau}=\p_x^2 \log (\tau \tilde \tau )+
\Bigl (\p_{x}\log \frac{\tilde \tau}{\tau}\Bigr )^2,
\eeq
or
\beq\label{sd2}
(D_2+D_1^2)\tau \cdot \tilde \tau =0,
\eeq
which is the mKP equation in the bilinear Hirota form. Let us put $t_3=0$ in this subsection
for simplicity. Then
$\tilde \tau = e^{zx+z^2 t_2}\hat \tau$, where $\hat \tau$ is the tau-function with
shifted times ${\bf t}=(t_1, t_2, t_3, \ldots )$: $\hat \tau ({\bf t})=
\tau ({\bf t}-[z^{-1}])$, where we have used the standard notation
$${\bf t}\pm [z^{-1}]=\Bigl (t_1\pm \frac{1}{z}, t_2 \pm 
\frac{1}{2z^2}, t_3 \pm \frac{1}{3z^3},
\, \ldots \Bigr ).$$ Equation (\ref{sd1}) becomes
\beq\label{sd3}
\p_{t_2}\log \frac{\hat \tau}{\tau}=\p_x^2 \log (\tau \hat \tau )+
\Bigl (\p_{x}\log \frac{\hat \tau}{\tau}\Bigr )^2 +2z \p_{x}\log \frac{\hat \tau}{\tau}.
\eeq
Let $y_i$ ($i=1, \ldots , N$) 
be zeros of the function $\hat \tau$. Then we can write
$$
\frac{\hat \tau}{\tau}=A e^{\alpha x +\beta t_2}\prod_i \frac{\sigma (x-y_i)}{\sigma (x-x_i)},
$$
where $A, \alpha, \beta$ are some constants. Substituting expressions for $\tau$,
$\hat \tau$ through $\sigma$-functions into (\ref{sd3}), we obtain:
$$
\sum_i \Bigl (\dot x_i \zeta (x-x_i)-\dot y_i \zeta (x-y_i)\Bigr )=
-\sum_i \Bigl (\wp (x-x_i)+\wp (x-y_i)\Bigr )
$$
$$
+\left (\sum_i \Bigl (\zeta (x-x_i)-\zeta (x-y_i)\Bigr )\right )^2+
\mu \sum_i \Bigl (\zeta (x-x_i)-\zeta (x-y_i)\Bigr ) +\mbox{const}\, ,
$$
where $\mu$ is a constant. Identifying residues at the poles at $x=x_i$ and $x=y_i$, we 
obtain the following system of first order differential equations:
\beq\label{sd4}
\left \{ \begin{array}{l}
\displaystyle{\dot x_i=2\sum_{j\neq i}\zeta (x_i-x_j)-2\sum_j \zeta (x_i-y_j)+\mu }
\\ \\
\displaystyle{\dot y_i=-2\sum_{j\neq i}\zeta (y_i-y_j)+2\sum_j \zeta (y_i-x_j)+\mu }.
\end{array}
\right.
\eeq
This is the so-called self-dual form of equations of motion of the elliptic 
Calogero-Moser system
\cite{ABW09,BSTV14}. For the first time it appeared in \cite{W82} as a B\"acklund
transformation of the Calogero-Moser system. 

It can be shown that equations (\ref{sd4}) are equivalent 
to (\ref{int1}). Indeed, let us differentiate the first equation in (\ref{sd4})
with respect to $t_2$:
\beq\label{sd5}
\begin{array}{lll}
\ddot x_i&=&\displaystyle{-2\sum_{j\neq i}(\dot x_i -\dot x_j)\wp (x_i-x_j)+
2\sum_j (\dot x_i -\dot y_j)\wp (x_i-y_j)}
\\ && \\
&=&\! \! -4\displaystyle{\sum_{j\neq i}\Bigl (\sum_{k\neq i}\zeta (x_i-x_k)\! -\! 
\sum_k \zeta (x_i-y_k)\! -\! \sum_{k\neq j}\zeta (x_j-x_k)\! +\! \sum_k \zeta (x_j -y_k)\Bigr )
\wp (x_i-x_j)}
\\ && \\
&& \! +4\displaystyle{\sum_{j}\Bigl (\sum_{k\neq i}\zeta (x_i-x_k)\! -\! 
\sum_k \zeta (x_i-y_k)\! +\! \sum_{k\neq j}\zeta (y_j-y_k)\! -\! \sum_k \zeta (y_j -x_k)\Bigr )
\wp (x_i-y_j)}.
\end{array}
\eeq
It can be proved \cite{BSTV14} (see the appendix) that the right hand side is in fact 
equal to $\displaystyle{4\sum_{j\neq i}\wp ' (x_i-x_j)}$. By symmetry, the same 
Calogero-Moser equations
of motion are satisfied by $y_i$'s. 

\subsection{Calogero-Moser system in discrete time}

The self-dual form of equations of motion is directly connected with the 
integrable time discretization of the Calogero-Moser system. To see this,
let us consider dynamics of poles of elliptic solutions to the semi-discrete
KP equation and
note that the disctete time flow $n$ in the KP hierarchy is introduced
according to the rule \cite{DJM82}
\beq\label{d1}
\tau^n ({\bf t})=\tau ({\bf t}-n[z^{-1}]),
\eeq
so that $\tau$ and $\hat \tau$ are tau-functions taken at two subsequent 
values of the discrete time. Accordingly, we can denote $x_i = x_i^n$,
$y_i=x_{i}^{n+1}$ and rewrite equations (\ref{sd4}) as
\beq\label{d2}
\left \{ \begin{array}{l}
\displaystyle{\dot x_i^n
=2\sum_{j\neq i}\zeta (x_i^n-x_j^n)-2\sum_j \zeta (x_i^n-x_j^{n+1})+\mu }
\\ \\
\displaystyle{\dot x_i^{n+1}=-2\sum_{j\neq i}\zeta (x_i^{n+1}
-x_j^{n+1})+2\sum_j \zeta (x_i^{n+1}-x_j^n)+\mu }.
\end{array}
\right.
\eeq
Shifting $n \to n-1$ in the second group of equations and subtracting the 
second line from the first one, we get equations of motion for the Calogero-Moser
system in discrete time \cite{NP94,NRK96}:
\beq\label{d3}
\sum_j \zeta (x_i^n-x_j^{n+1})+\sum_j \zeta (x_i^n-x_j^{n-1})
-2\sum_{j\neq i}\zeta (x_i^n-x_j^n)=0.
\eeq
Remarkably, these equations coincide with the nested Bethe ansatz equations
for the elliptic Gaudin model associated with the root system $A_m$, with the 
discrete time $n$ taking values $0, 1, \ldots , m+1$. 

\section{Elliptic solutions to the BKP equation}

\subsection{The BKP equation}

The BKP equation is the first member of an 
infinite BKP hierarchy with independent variables (``times'') 
$t_1=x$, $t_3, t_5, t_7, \ldots$ \cite{DJKM83,DJKM82}, see also \cite{DJKM82a,LW99,Tu07}. 
It is the following system of nonlinear partial differential
equations for two dependent variables $u$, $w$:
\beq\label{int2a}
\left \{
\begin{array}{l}
3w' =10 u_{t_3}+20 u^{'''} +120 uu'
\\ \\
w_{t_3}-6u_{t_5}=w^{'''}-6u^{{\scriptsize \rm V}}-60 uu^{'''}-60 u'u'' +6uw'-6wu',
\end{array}
\right.
\eeq
where prime means differentiation w.r.t. $x$. 
(Some notation in this section such as $w$
and $L,M$ below is the same as in the previous one but their meaning
is different; we hope that this will not lead to a misunderstanding.)
It is easy to see that the variable $w$ can be excluded 
and the equation can be written in terms a single dependent variable $U=\int^x udx$. 
Equations (\ref{int2}) are equivalent to the Zakharov-Shabat equation 
$\p_{t_5}B_3-\p_{t_3}B_5+[B_3, B_5]=0$ for the differential operators
\beq\label{int3}
B_3=\p_x^3+6u\p_x, \qquad B_5=\p_x^5+10u\p_x^3+10u'\p_x^2+w\p_x.
\eeq
Similarly to the case discussed in the previous section, 
the Zakharov-Shabat equation is the compatibility condition for 
the auxiliary linear problems
$$
\p_{t_3}\psi =B_3\psi , \qquad \p_{t_5}\psi =B_5\psi
$$
for the wave function $\psi$ which depends on a spectral parameter $z$.

The tau-function 
$\tau = \tau (t_1, t_3, t_5, \ldots )$ of the BKP hierarchy is related to the 
variables $u,w$ by the formulas
\beq\label{int4}
u=\p_x^2\log \tau , \qquad w=\frac{10}{3}\, \p_{t_3}\p_x \log \tau +
\frac{20}{3}\, \p_x^4 \log \tau +20 (\p_x^2 \log \tau )^2
\eeq
In terms of the tau-function, equations (\ref{int2a}) acquire the bilinear form \cite{DJKM82}
\beq\label{int5}
\Bigl (D_1^6 -5D_1^3D_3-5D_3^2+9D_1D_5 \Bigr ) \tau \cdot \tau =0.
\eeq
The wave function $\psi$
can be expressed through the tau-function according to the formula \cite{DJKM82}
\beq\label{psib}
\psi = A(z)\exp \, \Bigl (\, \sum_{k\geq 1, \, k \,\, {\rm odd}}t_k z^k \Bigr )\,
\frac{\tau \Bigr (t_1 -2z^{-1}, t_3-\frac{2}{3}\, z^{-3}, t_5-\frac{2}{5}\, z^{-5},
\ldots \Bigr )}{\tau (t_1, t_3, t_5, \ldots )},
\eeq
where $A(z)$ is a normalization factor.

\subsection{Dynamics of poles}

We are going to study dynamics of poles of elliptic in the variable $t_1=x$ solutions 
of the BKP equation as functions of $t_3=t$. For such solutions the tau-function  
has the same form (\ref{ell1}):
\beq\label{int6}
\tau = A e^{Q(x, t, \ldots )}\prod_{i=1}^{N}\sigma (x-x_i)
\eeq
with a quadratic form $Q=cx^2 +\ldots$ and $u$ is given by (\ref{ell2}).
The basic tool for 
studying $t$-dynamics of poles is the linear problem 
$\p_{t}\psi =B_3\psi$ for the 
function $\psi$, i.e.,
\beq\label{ba0}
\p_t \psi =\p_x^3\psi +6u \p_x \psi .
\eeq
As in the KP case,  
one can find double-Bloch solutions $\psi (x)$
\beq\label{ba1}
\psi = e^{xz+tz^3}\sum_{i=1}^N c_i \Phi (x-x_i, \lambda )
\eeq
with simple poles at $x=x_i$ and $x$-independent 
coefficients $c_i$ (the function $\Phi$ is the same as in the 
previous section).

It is evident from (\ref{ba0}) that the constant $c$ in 
the expression (\ref{ell2})
for the function $u$ can be eliminated
by the 
transformation $x\to x-12ct$, $t\to t$ (or $\p_x \to \p_x$, $\p_t \to \p_t +12c\p_x$
for the vector fields). Because of this we put $c=0$ from now on for simplicity.

Substituting (\ref{ba1}) into (\ref{ba0}) with  
$\displaystyle{u=-\sum_{i}\wp (x-x_i)}$, 
we get:
$$
\sum_i \dot c_i\Phi (x-x_i)-\sum_i c_i \dot x_i \Phi '(x-x_i)=3z^2
\sum_i c_i \Phi '(x-x_i)+3z\sum_i c_i \Phi ''(x-x_i)+\sum_i c_i \Phi '''(x-x_i)
$$
$$
-6z\Bigl (\sum_k \wp (x-x_k)\Bigr ) \Bigl (\sum_i c_i \Phi (x-x_i)\Bigr )
-6\Bigl (\sum_k \wp (x-x_k)\Bigr ) \Bigl (\sum_i c_i \Phi ' (x-x_i)\Bigr ).
$$
This expression has poles at $x=x_i$ (up to fourth order).
Poles of the fourth and third order cancel identically. 
As it can be seen 
by a direct calculation, the conditions of cancellation of second and first 
order poles have the form
\beq\label{ba2}
c_i\dot x_i=-(3z^2-3\wp (\lambda ))c_i -6z \sum_{k\neq i}c_k \Phi (x_i-x_k)-6
\sum_{k\neq i}c_k \Phi '(x_i-x_k)+6c_i \sum_{k\neq i}\wp (x_i-x_k),
\eeq
\beq\label{ba3}
\begin{array}{lll}
\dot c_i &=&\displaystyle{
3z\wp (\lambda ) c_i +2\wp '(\lambda )c_i-6z\sum_{k\neq i}c_k \Phi '(x_i-x_k)
-6zc_i\sum_{k\neq i}\wp (x_i-x_k)}
\\ &&\\
&&\displaystyle{-\, 6 \sum_{k\neq i}c_k \Phi ''(x_i-x_k)
+6c_i \sum_{k\neq i}\wp '(x_i-x_k)}
\end{array}
\eeq
which have to be valid for all $i=1, \ldots , N$.
In the matrix form, these conditions look like
linear problems for a vector ${\bf c} =(c_1, \ldots , c_N)^T$:
\beq\label{a1}
\left \{ \begin{array}{l}
L{\bf c} = 3(z^2 -\wp (\lambda )){\bf c}
\\ \\
\dot {\bf c} =M{\bf c},
\end{array}
\right.
\eeq
where
\beq\label{a1a}
L=-\dot X -6zA -6B +6D,
\eeq
\beq\label{a1b}
M= (3z\wp (\lambda ) +2\wp ' (\lambda ))I -6zB -6zD -6C +6D'
\eeq
with the same matrices
$X$, $A$, $B$, $C$, $D$, $D'$ as in the previous section.
Note that in the present case the matrices $L,M$ depend not only on $\lambda$
but also on $z$.
The compatibility condition of the linear problems (\ref{a1}) is 
\beq\label{a2}
\Bigl (\dot L+[L,M]\Bigr ) {\bf c} =0.
\eeq

In the appendix we prove
the following matrix identity:
\beq\label{a3}
\dot L+[L,M]=-12 D'\Bigl (L-3(z^2-\wp (\lambda ))I\Bigr )-\ddot X +
12D'(6D-\dot X)+6\dot D -6D''',
\eeq
where $\displaystyle{D'''_{ik}=\delta_{ik}\sum_{j\neq i}\wp '''(x_i-x_j)}$.
Using this identity, it is straightforward to see that
the compatibility condition (\ref{a2}) is equivalent 
to vanishing of all elements of the diagonal matrix
$
-\ddot X +12D'(6D-\dot X)+6\dot D -6D'''.
$
Writing the diagonal elements
explicitly, we get
equations of motion for the poles $x_i$:
$$
\ddot x_i +6\sum_{j\neq i}(\dot x_i +\dot x_j)\wp '(x_i-x_j)-72
\sum_{j\neq i}\sum_{k\neq i} \wp (x_i-x_j)\wp '(x_i-x_k)+6\sum_{j\neq i}
\wp '''(x_i-x_j)=0.
$$
Using the identity $\wp '''(x)=12\wp (x)\wp '(x)$, we get the
equations of motion (\ref{int8}). They were obtained in \cite{RZ18}. Their rational limit 
(when both periods tend to infinity and $\wp (x) \to 1/x^2$) is
\beq\label{a4}
\ddot x_i -12 \sum_{j\neq i}\frac{\dot x_i+\dot x_j}{(x_i-x_j)^3}+
144 \! \sum_{j\neq k\neq i}\frac{1}{(x_i-x_j)^2(x_i-x_k)^3}=0.
\eeq
We see that in contrast to the 
equations of motion for poles of elliptic solutions to the KP equation, where
interaction between ``particles'' (poles) is pairwise, in the BKP case
there is a three-body interaction.

\subsection{Integrals of motion}

Equations (\ref{int8}) do not admit a representation of Lax type. Instead,
the matrix relation
\beq\label{a5}
\dot L+[L,M]=-12 D' (L-\Lambda I),
\eeq
where $\Lambda =3(z^2-\wp (\lambda ))$, is
equivalent to the equations of motion (\ref{int8}). This is a sort of Manakov's triple 
representation. 
In contrast to the KP case, the evolution
$L\to L(t)$ of our ``Lax matrix'' 
is not isospectral. Nevertheless, the characteristic polynomial, 
$\det (L -\Lambda I)$,
is an integral of motion. Indeed, 
$$
\frac{d}{dt}\, \log \det (L-\Lambda I)=
\frac{d}{dt}\, \mbox{tr}\log (L-\Lambda I)
$$
$$
=\, \mbox{tr}\Bigl ( \dot L(L-\Lambda I)^{-1}\Bigr )=
-12\, \mbox{tr}D' =0,
$$
where we have used (\ref{a5}) and the fact that the matrix $D'$ is traceless
(because $\wp '$ is an odd function and so 
$\displaystyle{\sum_{i\neq j}\wp '(x_i-x_j)=0}$).
The expression 
$$
R(z, \lambda )=\det \Bigl (3(z^2 -\wp (\lambda ))I-L\Bigr )
$$ 
is a polynomial in $z$
of degree $2N$. Its coefficients are integrals of motion (some of them may be trivial). 

Applying the same similarity transformation $L=G\tilde L G^{-1}$ as in the previous section
with the matrix $G_{ij}=\delta_{ij}
e^{-\zeta (\lambda )x_i}$,
we conclude that the coefficients $R_k(\lambda )$ of the polynomial
\beq\label{a5a}
R(z, \lambda )=\sum_{k=0}^{2N}R_k(\lambda )z^k
\eeq
are elliptic functions of $\lambda$ with poles
at $\lambda =0$.

\noindent
{\it Example ($N=2$):}
$$
\det_{2\times 2} \Bigl (3(z^2 -\wp (\lambda ))I-L\Bigr )
=9z^4+3z^2\Bigl (\dot x_1 +\dot x_2 -18\wp (\lambda )\Bigr )-36 z\wp '(\lambda )
-3\wp (\lambda )(\dot x_1 +\dot x_2)
$$
$$\phantom{aaaaaaaaa}
+\dot x_1 \dot x_2 -6(\dot x_1 +\dot x_2)\wp (x_1-x_2)-27 \wp ^2(\lambda )+9g_2,
$$
where $g_2$ is the coefficient in the expansion of the $\wp$-function near $x=0$:
$\wp (x)=x^{-2}+\frac{1}{20}\, g_2 x^2 +\frac{1}{28}\, g_3 x^4 +O(x^6)$. 
Therefore, for $N=2$ there are 
two integrals of motion: $I_1=\dot x_1 +\dot x_2$, $I_2=
\frac{1}{2}(\dot x_1^2 + \dot x_2^2) +6(\dot x_1 +\dot x_2)\wp (x_1-x_2)$. 

\noindent
{\it Example ($N=3$):}
$$
\det_{3\times 3} \Bigl (3(z^2 -\wp (\lambda ))I-L\Bigr )=
27z^6+9\Bigl (I_1-45\wp (\lambda )\Bigr )z^4
-540\wp '(\lambda )z^3
$$
$$
+\left [ \frac{3}{2}\, I_1^2-3I_2-54\wp (\lambda )I_1
-1215\wp ^2(\lambda )+243g_2\right ]z^2
-36\wp '(\lambda )\Bigl (I_1+9\wp (\lambda )\Bigr )z
$$
$$
+I_3-I_1I_2 +\frac{1}{6}I_1^3 +3\wp (\lambda )\Bigl (I_2 -\frac{1}{2}I_1^2\Bigr )
-27\wp ^2 (\lambda )I_1
+9g_2 I_1 -135 \wp ^3 (\lambda )-27 g_2 \wp (\lambda ) +216 g_3,
$$
where
\beq\label{a6}
\begin{array}{lll}
I_1&=&\dot x_1 +\dot x_2 +\dot x_3,
\\ &&\\
I_2&=&\frac{1}{2}\Bigl (\dot x_1^2+\dot x_2^2+\dot x_3^2\Bigr )
+6\dot x_1 \Bigl (\wp (x_{12})+\wp (x_{13})\Bigr )+
6\dot x_2 \Bigl (\wp (x_{21})+\wp (x_{23})\Bigr )
\\ && \\
&&
+\, 6\dot x_3 \Bigl (\wp (x_{31})+\wp (x_{32})\Bigr )
-36 \Bigl (\wp (x_{12})\wp (x_{13})+\wp (x_{12})\wp (x_{23})
+\wp (x_{13})\wp (x_{23})\Bigr ),
\\&&\\
I_3&=&\frac{1}{3}\Bigl (\dot x_1^3+\dot x_2^3+\dot x_3^3\Bigr )
+6\dot x_1^2 \Bigl (\wp (x_{12})+\wp (x_{13})\Bigr )+
6\dot x_2^2 \Bigl (\wp (x_{21})+\wp (x_{23})\Bigr )
\\&&\\
&&+\, 6\dot x_3^2 \Bigl (\wp (x_{31})+\wp (x_{32})\Bigr )
+12 \dot x_1\dot x_2 \wp (x_{12})+12 \dot x_1\dot x_3 \wp (x_{13})+
12 \dot x_2\dot x_3 \wp (x_{23})
\\ &&\\
&& -\, 864 \wp (x_{12})\wp (x_{13}) \wp (x_{23})
\end{array}
\eeq
are integrals of motion
(here $x_{ik}\equiv x_i-x_k$).

For arbitrary $N$, one 
can prove that the quantities 
\beq\label{a7}
\begin{array}{l}
\displaystyle{I_1=\sum_{i}\dot x_i},
\\ \\
\displaystyle{I_2=\frac{1}{2}\sum_i \dot x_i^2+6 \sum_{i\neq j}\dot x_i \wp (x_{ij})
-18 \! \! \sum_{i\neq j\neq k}\wp (x_{ij})\wp (x_{ik})}
\end{array}
\eeq
are integrals of motion.
In the expression for $I_2$ the last sum is taken over all triples of distinct
numbers $i,j,k$ from $1$ to $N$. 
The proof can be found in \cite{RZ18}. It is based on the following identities
for the $\wp$-function:
\beq\label{a71}
\sum_{i=1}^n \p_{x_i}\! \prod_{k=1, \neq i}^n \wp (x_i-x_k)=0, \quad n=2, 3, \ldots 
\eeq
(for the proof we need them at $n=3$ and $n=4$). The proof of (\ref{a71}) is standard.
The left hand side is an elliptic function of $x_1$. Expanding it near possible poles 
at $x_1=x_k$, $k=2, \ldots , n$, one can see that it is regular, so it is a constant 
independent of $x_1$. By symmetry, this constant does not depend also on all the $x_i$'s.
To see that this constant is actually zero, one can put $x_k=kx$. 

Another integral of motion for arbitrary $N$ is 
\beq\label{a10}
J=\lim_{\lambda \to 0}R(\lambda ^{-1}, \lambda )=
\det_{1\leq i,j\leq N}\Bigl [\delta_{ij}\dot x_i -6\delta_{ij}
\sum_{k\neq i}\wp (x_{ik})-6(1-\delta_{ij})\wp (x_{ij})\Bigr ].
\eeq
Indeed, using the obvious formula
$\Phi '(x, \lambda )=\Phi (x, \lambda )
\Bigl (\zeta (x+\lambda )-\zeta (x)-\zeta (\lambda )\Bigr )$ and the expansion
$$
\tilde \Phi (x, \lambda )=e^{\zeta (\lambda )x}\Phi (x, \lambda )=
\lambda ^{-1}+\zeta (x)+\frac{1}{2}\, \frac{\sigma ''(x)}{\sigma (x)}\, \lambda +
O(\lambda ^2),
$$
we have
$$
\tilde L(z, \lambda )=(z-\lambda^{-1})Y(z, \lambda )+\dot X -6D -6Q +O(\lambda ),
$$
where $Q$ is the off-diagonal 
matrix with elements $Q_{ij}=(1-\delta_{ij})\wp (x_{ij})$
and $Y(z, \lambda )$ is a matrix which is regular at $z=\lambda^{-1}$. Therefore,
$R(\lambda^{-1}, \lambda )=\det (\dot X -6D -6Q) +O(\lambda )$.

\subsection{The spectral curve}

The spectral curve is given by the equation
\beq\label{s1}
R(z, \lambda )=\det \Bigl (3(z^2-\wp (\lambda ))I-L(z, \lambda )\Bigr )=0.
\eeq
It is easy to see that $L(-z, -\lambda )=L^{T}(z, \lambda )$, so the spectral curve 
admits the involution $\iota : (z, \lambda )\to (-z, -\lambda )$. 

As it was already mentioned, the coefficients $R_k(\lambda )$ in (\ref{a5a})
are elliptic functions of $\lambda$.
The functions 
$R_k (\lambda )$ obey the property $R_k(-\lambda )=(-1)^k R_k (\lambda )$
and
can be represented as linear combinations of $\wp$-function and
its derivatives. Coefficients
of this expansion are integrals of motion. Fixing values of these integrals, we obtain
via the equation $R(z, \lambda )=0$ the algebraic curve $\Gamma$ which is a 
$2N$-sheet covering of the initial elliptic curve ${\cal E}$ realized as a factor
of the complex plane with respect to the lattice generated by $2\omega$, $2\omega '$.

In a neighborhood of $\lambda =0$   
the matrix $\tilde L =G^{-1}L G$
can be written as
$$
\tilde L=-6\lambda ^{-1}(z-\lambda^{-1})(E-I)-6(z-\lambda^{-1})S+O(1),
$$ 
where
$E$ is the same rank $1$ matrix with matrix elements $E_{ij}=1$ as in the previous section
and $S$ is the antisymmetric matrix with matrix elements $S_{ij}=\zeta (x_i-x_j)$,
$i\neq j$, $S_{ii}=0$.
Therefore, near $\lambda =0$ the function $R(z, \lambda )$ can be represented in the form
$$
R(z, \lambda )=\det \Bigl (3(z^2-\lambda^{-2})I+6\lambda^{-1}(z-\lambda^{-1})(E-I)+6
(z-\lambda^{-1})S+O(1)\Bigr )
$$
$$
=\, \det \Bigl (3(z-\lambda^{-1})^2I +6\lambda^{-1}(z-\lambda^{-1})E
+6(z-\lambda^{-1})S+O(1)\Bigr )
$$
$$
=\, 3^N (z-\lambda^{-1})^{2N}\det \left (I+\frac{2}{z\lambda -1}\, E +
\frac{2\lambda}{z\lambda -1}\, S +O(\lambda^2)\right ).
$$
Using the fact that $\det \Bigl (A+\varepsilon B\Bigr )=\det A 
\Bigl (1+\varepsilon \, \mbox{tr}\, (A^{-1}B)\Bigr )+O(\varepsilon ^2)$ for any two
matrices $A$, $B$ and the relation 
$(I-\alpha E)^{-1}=I+\frac{\alpha}{1-N\alpha}\, E$, we find
$$
\det \left (I+\frac{2}{z\lambda -1}\, E +
\frac{2\lambda}{z\lambda -1}\, S +O(\lambda^2)\right )
$$
$$
=\, \det \left (I+\frac{2}{z\lambda -1}\, E+O(\lambda^2) \right )
\left (1+\frac{2\lambda}{z\lambda -1}\, \mbox{tr} \, \Bigl (S -
\frac{2}{z\lambda \! +\! 2N\! -\! 1}\, ES\Bigr )+O(\lambda^2) \right ).
$$
For any antisymmetric matrix $S$ we have 
$\mbox{tr}\, S =\mbox{tr}\, (ES)=0$, so we are left with
$$
R(z, \lambda )=3^N (z-\lambda^{-1})^{2N}\det \left (I+\frac{2}{z\lambda -1}\, E +
O(\lambda ^2)\right ).
$$
Therefore, we can write $R(z, \lambda )$ in the form
\beq\label{s2}
R(z, \lambda )=3^N \Bigl (z+(2N-1)\lambda^{-1}-f_{2N}(\lambda )\Bigr )
\Bigl (z-\lambda^{-1}-f_{1}(\lambda )\Bigr )
\prod_{i=2}^{2N-1}\Bigl (z-\lambda^{-1}-f_{i}(\lambda )\Bigr ),
\eeq
where $f_i$ are regular functions of $\lambda$ at $\lambda =0$. 
This means that the function $z$
has simple poles on all sheets at the points $P_j$ ($j=1, \ldots , 2N$) located 
above $\lambda =0$.
The involution $\iota$ implies that $f_{2N}$ and $f_1$ 
are odd functions: $f_{2N}(-\lambda )=-f_{2N}(\lambda )$, 
$f_{1}(-\lambda )=-f_{1}(\lambda )$.
The other sheets can be 
numbered in such a way that 
$f_i (-\lambda )=-f_{2N+1-i}(\lambda )$,
$i=2, 3, \ldots , N$.
We have the following expansions 
of the function $z$ near 
the ``points at infinity'' $P_j$:
\beq\label{s2a}
\begin{array}{l}
z=\, \lambda^{-1}+f_j(\lambda ) \quad \mbox{near $P_j$}, \quad j=1, \ldots , 2N-1,
\\ \\
z=-(2N\! -\! 1)\lambda ^{-1}+f_{2N}(\lambda ) \quad \mbox{near $P_{2N}$}.
\end{array}
\eeq
Similarly to the spectral curve of the elliptic
Calogero-Moser model (\ref{spec2}), one of the sheets is distinguished. 
We call it the upper sheet. There is also another distinguished sheet,
where the point $P_1$ is located.
We call it the lower sheet for brevity. The points $P_1, P_{2N}$ are two 
fixed points of the involution $\iota$.

The genus $g$ 
of the spectral curve $\Gamma$ can be found by an argument which
is similar to the one in section 2.5. We have $2g-2=\nu$, where $\nu$ is the number 
of ramification points of the covering $\Gamma \to {\cal E}$. 
The ramification points are zeros on $\Gamma$
of the function $\p R/\p z$. 
The function $\p R/\p z$ has simple poles at the points $P_j$ 
($j=1, \ldots , 2N-1$) on all
sheets except the upper one, where it has a pole of order $2N-1$. Therefore,
$\nu = 2(2N-1)$ and $g=2N$.

\subsection{Analytic properties of the $\psi$-function on the spectral curve}

Similarly to section \ref{section:ba}, the coefficients
$c_i$ in the pole ansatz for the function $\psi$
are functions on the spectral curve $\Gamma$:
$c_i=c_i(t, P)$ ($P=(z, \lambda )$ is a point on the curve).
Let us normalize them by the condition $c_1(0, P)=1$.
After normalization the components
$c_i(0,P)$ become meromorphic functions on $\Gamma$ outside the points $P_j$ located
above $\lambda =0$. The location of their poles depends on the initial data.


On all sheets except the lower one the leading term of the matrix $\tilde L$ 
as $\lambda \to 0$ is proportional to $E-I$. 
Finding explicitly eigenvectors of the matrix $E-I$, one can see that
near the ``points at infinity'' $P_j$ ($j=2, \ldots , 2N$) the functions 
$c_i (0, P)$ have the form
\beq\label{s3}
c_i (0, P)=\Bigl (c_i^{0(j)} +O(\lambda )\Bigr )e^{-\zeta (\lambda )(x_i(0)-x_1(0))},
\quad 2\leq i\leq N, \quad j=2, \ldots 2N-1
\eeq
on all sheets except the lower and upper ones. Here
$\displaystyle{\sum_{i=2}^{N}c_i^{0(j)}=-1}$. On the upper sheet
\beq\label{s4}
c_i (0, P)=\Bigl (1 +O(\lambda )\Bigr )e^{-\zeta (\lambda )(x_i(0)-x_1(0))},
\quad 2\leq i\leq N, \quad j= 2N.
\eeq
On the lower sheet, the leading term of the matrix $\tilde L$ 
as $\lambda \to 0$ is $O(1)$. Expanding the matrix $\tilde L$ in powers of $\lambda$, we
have
$$
\Lambda I-\tilde L=6f_1'(0)E+\dot X -6D -6Q +O(\lambda ),
$$
where $Q$ is the matrix with matrix elements $Q_{ij}=(1-\delta_{ij})\wp (x_i-x_j)$.
Let $c_i^{0(1)}$ be the eigenvector of the matrix $6f_1'(0)E+\dot X -6D -6Q$ 
(taken at $t=0$) with zero eigenvalue normalized by the condition $c_1^{0(1)}=1$,
then in a neighborhood of the point $P_1$ we have
\beq\label{s3a}
c_i (0, P)=\Bigl (c_i^{0(1)} +O(\lambda )\Bigr )e^{-\zeta (\lambda )(x_i(0)-x_1(0))},
\quad 2\leq i\leq N, \quad j=1.
\eeq

Let ${\cal S}(t)$ be 
the fundamental matrix of solutions to the equation 
$\p_t {\cal S} =M{\cal S}$, ${\cal S}(0)=I$. It is a regular function of 
$z, \lambda$ for $\lambda \neq 0$.
Using the Manakov's triple representation (\ref{a5}), we can write
$$
\Bigl (\dot L +[L,M]+12D' (L-\Lambda I)\Bigr ){\bf c}(t)=0,
\quad \Lambda = 3(z^2 -\wp (\lambda )).
$$
Using the relations ${\bf c}(t)={\cal S}(t){\bf c}(0)$ and $M=\dot {\cal S} {\cal S}^{-1}$, 
we rewrite this equation
as
$$
\Bigl [\p_t \Bigl ({\cal S}^{-1}(L-\Lambda I){\cal S}\Bigr ) +
12{\cal S}^{-1}D' (L-\Lambda I){\cal S}\Bigr ]{\bf c}(0)=0.
$$
Equivalently, we can represent it in the form of the differential equation
$$
\p_t {\bf b}(t)=W(t){\bf b}(t), \qquad W(t)=12{\cal S}^{-1}D'{\cal S},
$$
for the vector
$
{\bf b}(t)={\cal S}^{-1}(L-\Lambda I){\bf c}(t)$ with the initial condition ${\bf b}(0)=0$.
The differential equation with zero initial condition has the unique solution
${\bf b}(t)=0$ for all $t>0$.
It then follows that ${\bf c}(t)={\cal S}(t){\bf c}(0)$ is the 
common solution of the equations $\dot {\bf c}=M{\bf c}$ and $L{\bf c}=\Lambda {\bf c}$
for all $t>0$.
Therefore, the vector ${\bf c}(t, P)$ has the same $t$-independent poles as 
${\bf c}(0,P)$.

Similarly to section \ref{section:ba}, the next step is to pass to the gauge equivalent pair 
$\tilde L$, $\tilde M$, where
$$
\tilde L=G^{-1}LG , \quad \tilde M =-G^{-1}\p_t G +G^{-1}MG
$$
with the same diagonal matrix $G_{ij}=\delta_{ij}e^{-\zeta (\lambda )x_i}$ as before. 
The gauge-transformed linear system is
$$
\tilde L \tilde {\bf c}=3(z^2\! -\! \wp (\lambda ))\tilde {\bf c}, \qquad
\p_t \tilde {\bf c}=\tilde M \tilde {\bf c}, 
$$
where $\tilde {\bf c}=G^{-1}{\bf c}$,  ${\bf c}=(c_1, \ldots , c_N)^T$.

It is a straightforward calculation to verify that the following relation holds:
\beq\label{s5}
\tilde M=-\lambda^{-1}\tilde L+(3z\lambda^{-2}\! -\! 4\lambda^{-3})I+
6(z-\lambda^{-1})(Q-D) +O(1).
\eeq
(It should be taken into account that $z$ is of order $O(\lambda^{-1})$, see (\ref{s2a}),
so the terms proportional to $z$ have to be kept in the expansion.)
Applying the both sides to the eigenvector $\tilde {\bf c}$ of $\tilde L$
with the eigenvalue $\Lambda =
3(z^2\! -\! \wp (\lambda ))=3(z^2-\lambda^{-2})+O(\lambda^2)$, we get
\beq\label{s6}
\p_t \tilde {\bf c}=-z^3\tilde {\bf c}+(z-\lambda^{-1})^3\tilde {\bf c}
+6(z-\lambda^{-1})(Q-D)\tilde {\bf c} +O(1).
\eeq
Therefore, since $z=\lambda^{-1}+O(1)$ on all sheets except the upper one, we have
\beq\label{s7}
\p_t \tilde {\bf c}^{(j)}=-(z^3+O(1))\tilde {\bf c}^{(j)}, \quad j=1, \ldots , 2N-1,
\eeq
so 
$$
\tilde {\bf c}^{(j)}(t,P)=({\bf c}^{0(j)}+O(\lambda ))
e^{-z^3t}, \quad j=1, \ldots , 2N-1.
$$
On the
upper sheet, the corresponding eigenvector $\tilde {\bf c}^{2N}$ of the matrix $\tilde L$
is proportional to the vector ${\bf e}=(1, 1, \ldots , 1)^T$  
(plus terms of order $O(1)$). Note that $(Q-D){\bf e}=0$.
Therefore, since $z=-(2N-1)\lambda^{-1}+f_{2N}$ on the upper sheet, we have
\beq\label{s7a}
\p_t \tilde {\bf c}^{(2N)}=\Bigl 
(-z^3+k^3(\lambda ) +O(1)\Bigr )\tilde {\bf c}^{(2N)}, 
\eeq
where
$$
k(\lambda )=-2N\lambda^{-1}+f_{2N},
$$
so
$$
\tilde {\bf c}^{(2N)}(t,P)=({\bf e}+O(\lambda ))e^{(-z^3+k^3(\lambda ))t} .
$$
Hence the normalized vector ${\bf c}(t,P)$ is of the form
\beq\label{s8}
c_i^{(j)}(t,P)=c_{ij}(\lambda ) e^{-\zeta (\lambda )(x_i(t)-x_1(0))+\nu_j (\lambda )t},
\eeq
where $\nu_j =-z^3$ for $j=1, \ldots , 2N-1$, $\nu_{2N}=-z^3 +k^3 (\lambda )$ and
$c_{ij}(\lambda )$ are regular functions in a neighborhood of $\lambda =0$. Their
values at $\lambda =0$ are
\beq\label{s9}
c_{1j}(0)=1, \quad j=1, \ldots , 2N, \qquad c_{ij}(0)=c_i^{0(j)}, \quad i\geq 2, \,\,
j\neq 2N, \qquad c_{i\, 2N}(0)=1,
\eeq
with $\displaystyle{\sum_{i=2}^{N}c_i^{0(j)}=-1}$ for $j=2, \ldots , 2N-1$.

Analytic properties of the function 
$$
\psi (x, t, P)=\sum_{i=1}^{N}c_i(t, P)\Phi (x-x_i , \lambda )e^{zx+z^3t}
$$
follow from those of the vector ${\bf c}(t,P)$.
Equation (\ref{s8}) implies that in the function
$\psi$ the essential 
singularities at $\lambda =0$ cancel on all sheets except the upper one, where 
$\psi \propto e^{k(\lambda )x+k^3(\lambda )t}e^{\zeta (\lambda )x_1(0)}$. 
From (\ref{s9}) we see that
$\psi$ has simple poles at the points $P_1, P_{2N}$ (the two fixed points
of the involution $\iota$) and no poles at the points 
$P_j$ for $j=2, \ldots 2N-1$. The residue at the pole at $P_1$ is constant 
as a function of $x, t$. This is in agreement with the fact that the differential
operators $B_3, B_5$ (\ref{int3}) have no free terms, and so the result of 
their action to a 
constant vanishes.

The function $\psi$ also has other poles in the
finite part of the curve $\Gamma$, which do not depend on $x,t$.

\subsection{Self-dual form of the equations of motion}

Let us represent the wave function in (\ref{ba0}) as $\psi =\tilde \tau / \tau$, then 
the linear problem acquires the form
\beq\label{sdb1}
\p_{t_3}\log \frac{\tilde \tau}{\tau}=\p_x^3 \log  \frac{\tilde \tau}{\tau}+
+3\p_{x}\! \log \frac{\tilde \tau}{\tau}\,\, \p_x^2\log (\tau \tilde \tau )+
\Bigl (\p_{x}\log \frac{\tilde \tau}{\tau}\Bigr )^3,
\eeq
or
\beq\label{sdb2}
(D_3-D_1^3)\tau \cdot \tilde \tau =0.
\eeq
By analogy with the KP case, this equation can be called the modified BKP equation. 
We have
$\tilde \tau = e^{zx+z^3 t_3}\hat \tau$, where
$$
\hat \tau (t_1, t_3, t_5, \ldots )=\tau \Bigl (t_1-\frac{2}{z},
t_3-\frac{2}{3z^3}, t_5-\frac{2}{5z^5}, \ldots \Bigr ).
$$
(see (\ref{psib})). Equation (\ref{sdb1}) becomes
\beq\label{sdb3}
\begin{array}{c}
\displaystyle{
\p_{t_3}\log \frac{\hat \tau}{\tau}=\p_x^3 \log  \frac{\hat \tau}{\tau}
+3\p_{x}\! \log \frac{\hat \tau}{\tau}\,\, \p_x^2\log (\tau \hat \tau )+
\Bigl (\p_{x}\log \frac{\hat \tau}{\tau}\Bigr )^3}
\\ \\
\displaystyle{+\, 3z^2 \p_x \log \frac{\hat \tau}{\tau}+3z\p_x^2\log (\tau \hat \tau )
+3z \Bigl (\p_{x}\log \frac{\hat \tau}{\tau}\Bigr )^2.}
\end{array}
\eeq
We can write
$$
\frac{\hat \tau}{\tau}=Ae^{\alpha x +\beta t}\prod_i \frac{\sigma (x-y_i)}{\sigma (x-x_i)},
$$
where $A, \alpha , \beta$ are constants and $y_i$ are zeros of the function $\hat \tau$.
Identifying (first order) poles in (\ref{sdb3}) at $x_i$ and $y_i$, we obtain
\beq\label{sdb4}
\left \{
\begin{array}{lll}
\dot x_i&=& \displaystyle{3\sum_{j\neq i}\wp (x_i-x_j)+3\sum_j \wp (x_i-y_j)-
3\Bigl (\sum_{j\neq i}\zeta (x_i-x_j)\! -\! \sum_j \zeta (x_i-y_j)\Bigr )^2}
\\ &&\\
&&+\, 
6\mu \displaystyle{\sum_{j\neq i}\zeta (x_i-x_j)-6\mu \sum_j \zeta (x_i-y_j)-3\mu^2}
\\ &&\\
\dot y_i&=& \displaystyle{3\sum_{j\neq i}\wp (y_i-y_j)+3\sum_j \wp (y_i-x_j)-
3\Bigl (\sum_{j\neq i}\zeta (y_i-y_j)\! -\! \sum_j \zeta (y_i-x_j)\Bigr )^2}
\\ &&\\
&&-\, 
6\mu \displaystyle{\sum_{j\neq i}\zeta (y_i-y_j)+6\mu \sum_j \zeta (y_i-x_j)-3\mu^2},
\end{array}\right.
\eeq
where $\mu$ is a constant. This is the self-dual form of equations of motion 
(\ref{int8}).  A direct verification of the equations of motion for $x_i$ 
using calculation of
$\ddot x_i$ from (\ref{sdb4}) is too complicated to be made explicitly.

\subsection{Dynamics of poles of elliptic solutions to the No\-vi\-kov-\-Ve\-se\-lov
equation}

The Novikov-Veselov equation \cite{NV84} is a close relative of the BKP equation. 
In fact it is a member of the 2-component BKP hierarchy. The equation reads as
\beq\label{nv1}
v_t = v_{xxx}+v_{\bar x \bar x \bar x}+6(uv)_x +6(\bar u v)_{\bar x}
\eeq
with the additional constraints
\beq\label{nv2}
u_{\bar x}=v_x, \qquad \bar u_x=v_{\bar x}.
\eeq
These equations are compatibility conditions of the linear problems
\beq\label{nv3}
(\p_x \p_{\bar x}+2v)\psi =0,
\eeq
\beq\label{nv4}
\p_t \psi =(\p_x^3 +\p_{\bar x}^3 +6u\p_x +6\bar u \p _{\bar x})\psi
\eeq
for the wave function $\psi$. The tau-function is connected with $u, \bar u, v$ by the
formulas
\beq\label{nv5}
v=\p_x \p_{\bar x}\log \tau , \quad
u=\p_x^2 \log \tau , \quad
\bar u = \p_{\bar x}^2 \log \tau .
\eeq
For the solutions $u, \bar u, v$ that are elliptic functions of $x$ we have
\beq\label{nv6}
\tau = Ce^{\gamma x\bar x}\prod_{i=1}^{N}\sigma (x-x_i),
\eeq
where $\gamma$ is a constant. The roots $x_i$ depend on $\bar x$ and $t$.

We will be interested in the dynamics of the $x_i$'s as functions of $\bar x$.
Substituting the pole ansatz for the wave function
\beq\label{nv7}
\psi = e^{xz+\bar x z^{-1}}\sum_i c_i \Phi (x-x_i, \lambda )
\eeq
into (\ref{nv3}), we have:
$$
(1+2\gamma )\sum_i c_i \Phi (x-x_i)-z\sum_i c_i \dot x_i \Phi ' (x-x_i)
+z\sum_i \dot c_i \Phi (x-x_i)+\sum_i \dot c_i \Phi '(x-x_i)
$$
$$
+z^{-1}\sum_i c_i \Phi ' (x-x_i)-\sum_i c_i \dot x_i\Phi '' (x-x_i)
+2\Bigl (\sum_i \dot x_i \wp (x-x_i)\Bigr )\Bigl (\sum_k c_k\Phi (x-x_k)\Bigr )
=0,
$$
where dot denotes the $\bar x$-derivative (in this subsection only). The cancellation
of poles leads to the following conditions:
$$
(1+2\gamma )c_i+2\dot x_i \sum_{k\neq i}c_k \Phi ' (x-x_k)+2c_i \sum_{k\neq i}
\dot x_k \wp (x_i-x_k)-\wp (\lambda )\dot x_i c_i +z\dot c_i=0,
$$
$$
z\dot x_i c_i -z^{-1}c_i +2\dot x_i \sum_{k\neq i}c_k \Phi (x-x_k)-\dot c_i =0.
$$
They can be rewritten in the matrix form as
\beq\label{nv8}
\left \{
\begin{array}{l}
L{\bf c}=3(z^2 \! -\! \wp (\lambda )){\bf c}
\\ \\
\dot {\bf c}=\hat M {\bf c},
\end{array}
\right.
\eeq
where
$$
L=-6\dot X^{-1}({\cal D} +\gamma I)-6zA -6B,
$$
$$
\hat M=-z^{-1}I+z\dot X +2\dot XA
$$
and $\displaystyle{{\cal D}_{ij}=\delta_{ij}\sum_{k\neq i}\dot x_k \wp (x_i-x_k)}$.
In fact this matrix $L$ should be the same as the matrix $L$ in (\ref{a1a}).
Their off-diagonal parts coincide as written. Equating their diagonal parts,
we get the following 
relations between velocities with respect to $t_3$ and $\bar x$ (which
could be denoted as $\bar t_1$):
\beq\label{nv9}
\p_{\bar x}x_i \cdot \p_{t_3}x_i=6\sum_{j\neq i}
(\p_{\bar x}x_i + \p_{\bar x}x_j)\wp (x_i-x_j)+6\gamma .
\eeq

With the help of identities used in section 2.3 one can prove the following matrix identity:
\beq\label{nv10}
\dot L +[L, \hat M]=2[A, \dot X]\Bigl (L-3(z^2 \! -\! \wp (\lambda )\Bigr )+
6\dot X^{-2}\ddot X {\cal D}-6\dot X^{-1}\dot  {\cal D}+12{\cal D}',
\eeq
where $\displaystyle{{\cal D}'_{ij}=\delta_{ij}\sum_{k\neq i}\dot x_k \wp '(x_i-x_k)}$.
Therefore, the compatibility condition of the linear problems (\ref{nv8}) is vanishing
of the diagonal matrix $\dot X^{-2}\ddot X {\cal D}-\dot X^{-1}\dot  {\cal D}+2{\cal D}'$, which
leads to the equations of motion
\beq\label{nv11}
\sum_{j\neq i}(\dot x_i \ddot x_j-\ddot x_i \dot x_j)\wp (x_i-x_j)-
\sum_{j\neq i}\dot x_i \dot x_j (\dot x_i +\dot x_j)\wp '(x_i-x_j)=0.
\eeq
They are equivalent to the Manakov's triple representation
\beq\label{nv12}
\dot L +[L, \hat M]=2[A, \dot X] (L-\Lambda I), \qquad \Lambda =3(z^2 \! -\! \wp (\lambda )).
\eeq
The corresponding spectral curve is the same as for the BKP equation.

The rational degeneration of equations (\ref{nv11}) reads
\beq\label{nv13}
\sum_{j\neq i}\frac{\dot x_i \ddot x_j-\ddot x_i \dot x_j}{(x_i-x_j)^2}+2
\sum_{j\neq i}\frac{\dot x_i \dot x_j (\dot x_i +\dot x_j)}{(x_i-x_j)^3}=0.
\eeq
It was noticed by A.Zotov (see the remark in \cite{Krichever06}) that these equations 
can be resolved with respect to the $\ddot x_j$'s and are equivalent to
\beq\label{nv14}
\ddot x_i=2\sum_{j\neq i}\frac{\dot x_i \dot x_j}{x_i-x_j},
\eeq
which can be regarded as a limiting case of the rational Ruijsenaars-Schneider system.
However, the elliptic system (\ref{nv11}) hardly admits such a simple resolution. 

\section{Elliptic solutions to the 2D Toda equation}

\subsection{The 2D Toda equation}

The 2D Toda equation is the first member of the infinite
2DTL hierarchy \cite{UT84}. It is equivalent to the zero curvature equation
$\p_{\bar t_1}C_1 -\p_{t_1}\bar C_1 +[C_1, \bar C_1]=0$ for the 
difference operators
\beq\label{t1}
C_1=e^{\eta \p_x}+b(x), \qquad \bar C_1 =a(x)e^{-\eta \p_x},
\eeq
which, in its turn, is the compatibility condition of the linear problems
\beq\label{t2}
\begin{array}{l}
\p_{t_1}\psi (x)=\psi (x+\eta )+b(x)\psi (x),
\\ \\
\p_{\bar t_1}\psi (x)=a(x)\psi (x-\eta ).
\end{array}
\eeq
Writing it explicitly, we obtain the system
$$
\left \{ \begin{array}{l}
\p_{t_1}\log a(x)=b(x)-b(x-\eta )
\\ \\
\p_{\bar t_1}b(x)=a(x)-a(x+\eta ).
\end{array}
\right.
$$
Excluding $b(x)$, we get the second order differential equation for $a(x)$
\beq\label{t3}
\p_{t_1}\p_{\bar t_1}\log a(x)=2a(x)-a(x+\eta )-a(x-\eta )
\eeq
which is one of the forms of the 2D Toda equation. In terms of the function
$\varphi (x)$ introduced through the relation $a(x)=e^{\varphi (x)-\varphi (x-\eta )}$
it acquires the most familiar form
\beq\label{t4}
\p_{t_1}\p_{\bar t_1}\varphi (x)=e^{\varphi (x)-\varphi (x-\eta )}-
e^{\varphi (x+\eta )-\varphi (x)}.
\eeq
The change of the dependent variables from $a, b$ to the tau-function,
\beq\label{t5}
a(x)=\frac{\tau (x+\eta )\tau (x-\eta )}{\tau^2 (x)}, \qquad
b(x)=\p_{t_1}\log \frac{\tau (x+\eta )}{\tau (x)}, 
\eeq
brings the 2D Toda equation to the bilinear form \cite{JM83}
\beq\label{t6}
\frac{1}{2}\, D_{1}D_{\bar 1}\tau (x)\cdot \tau (x)=\tau ^2(x)-\tau (x+\eta )\tau (x-\eta ),
\eeq
or
\beq\label{t7}
\p_{t_1}\p_{\bar t_1}\log \tau (x)=1-\frac{\tau (x+\eta )\tau (x-\eta )}{\tau^2 (x)}.
\eeq
The constant term in the right hand side is chosen from the condition that 
$\tau (x)=\mbox{const}$ be a solution.

\subsection{Dynamics of poles of elliptic solutions}

We are interested in solutions for which $a(x), b(x)$ are elliptic functions of the 
variable $x$. For such solutions the tau-function has the form
\beq\label{te1}
\tau (x)=Ce^{cx^2 +rxt_1 +\bar rx\bar t_1 +\gamma t_1 \bar t_1}\prod_{i=1}^{N}\sigma (x-x_i),
\eeq
then
$$
a(x)=e^{2\eta ^2c}\prod_k \frac{\sigma (x-x_k+\eta )\sigma (x-x_k-\eta )}{\sigma ^2(x-x_k)}
=e^{2\eta ^2c}\sigma ^{2N}(\eta )\prod_k \Bigl (\wp (\eta )-\wp (x-x_k)\Bigr ),
$$
$$
b(x)=\sum_k \dot x_k \Bigl (\zeta (x-x_k)-\zeta (x-x_k+\eta )\Bigr )+r\eta ,
$$
where $\dot x_k=\p_{t_1}x_k$. 

We begin with investigating the dynamics of poles
as functions of the time $t_1$. 
To this end, 
it is enough to solve the first linear problem in 
(\ref{t2}) with $b(x)$ as above and the following pole ansatz for the $\psi$-function:
\beq\label{te2}
\psi =z^{x/\eta}e^{t_1z+\bar t_1z^{-1}}\sum_{i=1}^N c_i \Phi (x-x_i, \lambda ).
\eeq
Note that in \cite{KZ95} a slightly different function $\Phi$ was used, which 
differs from the present one by the exponential factor. Substituting (\ref{te2}) into
(\ref{t2}), we get:
$$
z\sum_i c_i \Phi (x-x_i)+\sum_i \dot c_i \Phi (x-x_i)-\sum_i c_i \dot x_i \Phi '(x-x_i)=
z\sum_i c_i \Phi (x-x_i+\eta )
$$
$$+\left (
\sum_k \dot x_k \Bigl (\zeta (x-x_k)-\zeta (x-x_k+\eta )\Bigr )+r\eta \right )
\sum_i c_i \Phi (x-x_i).
$$
The cancellation of poles leads to the conditions
$$
\left \{ \begin{array}{l}
\displaystyle{zc_i+\dot c_i=r\eta c_i +\dot x_i \sum_{k\neq i}c_k \Phi (x-x_k)
+c_i\sum_{k\neq i}\dot x_k \Bigl (\zeta (x_i-x_k)-\zeta (x_i-x_k+\eta )\Bigr )}
\\ \\
\displaystyle{zc_i-\dot x_i \sum_k c_k \Phi (x_i-x_k-\eta )=0},
\end{array}
\right.
$$
which can be written in matrix form
\beq\label{te3}
\left \{ \begin{array}{l}
L{\bf c}=z{\bf c}
\\ \\
\dot {\bf c}=M{\bf c}
\end{array}
\right.
\eeq
with the matrices
$
L=\dot X A^{-}$, $M=r\eta I+\dot X A -\dot X A^{-}+D^{0}-D^{+},
$
where $A$ is the same matrix as before and
$$
A^{-}_{ij}=\Phi (x_i-x_j-\eta ), \quad
D^{\pm}_{ij}=\delta_{ij}\sum_{k\neq i}\dot x_k \zeta (x_i-x_k\pm \eta ), 
\quad
D^{0}_{ij}=\delta_{ij}\sum_{k\neq i}\dot x_k \zeta (x_i-x_k ).
$$
A direct calculation shows that
$$
\dot L+[L,M]=\Bigl (\ddot X \dot X^{-1}+D^++D^--2D^0\Bigr )L,
$$
so the compatibility condition of the linear problems (\ref{te3}) is
$\ddot X \dot X^{-1}\! +\! D^+ \! +\! D^- \! -\! 2D^0=0$,
which implies equations of motion
\beq\label{te4}
\begin{array}{lll}
\ddot x_i &=&\displaystyle{-\sum_{k\neq i}\dot x_i\dot x_k \Bigl (
\zeta (x_i-x_k+\eta )+\zeta (x_i-x_k-\eta )-2\zeta (x_i-x_k)\Bigr )}
\\ && \\
&=&\displaystyle{\sum_{k\neq i}\dot x_i\dot x_k\frac{\wp '(x_i-x_k)}{\wp (\eta )-
\wp (x_i-x_k)}}
\end{array}
\eeq
together with their Lax representation. These are equations of motion for the 
elliptic Ruijsenaars-Schneider $N$-body system (a relativistic generalization of the 
Calogero-Moser system). 

It can be directly verified that the Ruijsenaars-Schneider system is Hamiltonian
with the Hamiltonian
\beq\label{te5}
H=\sum_i e^{p_i}\prod_{k\neq i}\frac{\sigma (x_i-x_k+\eta )}{\sigma (x_i-x_k)},
\eeq
where $p_i, x_i$ are canonical variables. Clearly,
\beq\label{te8a}
H=\sum_i \dot x_i =\mbox{const}\, \mbox{tr}\, L.
\eeq

Let us now investigate dynamics of poles as functions of $\bar t_1$.
Substitution of the pole ansatz into the second linear problem in (\ref{t2}) leads
to rather cumbersome calculations and necessity to use complicated identities for 
elliptic functions. Below we follow another way. Substituting the expression
(\ref{te1}) for the tau-function into equation (\ref{t7}), we get
$$
\gamma \! -\! 1 -\sum_i \p_{\bar t_1}\p_{t_1} x_i \zeta (x\! -\! x_i) \! -\! 
\sum_i \p_{\bar t_1}x_i \p_{t_1}x_i \wp (x\! -\! x_i)=\!
-e^{2c\eta ^2}\prod_i \frac{\sigma (x\! -\! x_i \! +\! \eta )
\sigma (x\! -\! x_i \! -\! \eta )}{\sigma^2 (x-x_i)}.
$$
Equating the coefficients at the first and second order poles, we obtain the 
relations
\beq\label{te6}
\p_{\bar t_1}x_i \, \p_{t_1}x_i=-e^{2c\eta ^2}\sigma^2(\eta )
\prod_{k\neq i}\frac{\sigma (x_i\! -\! x_k \! +\! \eta )
\sigma (x_i\! -\! x_k \! -\! \eta )}{\sigma^2 (x_i-x_k)},
\eeq
\beq\label{te7}
\p_{\bar t_1}\p_{t_1} x_i=\p_{\bar t_1}x_i \, \p_{t_1}x_i 
\sum_{k\neq i}\Bigl (\zeta (x_i-x_k+\eta )+\zeta (x_i-x_k-\eta )-2
\zeta (x_i-x_k )\Bigr )
\eeq
(they were mentioned in \cite{KZ95}). Let us differentiate logarithm of equation
(\ref{te6}) with respect to $\bar t_1$ and use (\ref{te7}). In this way we obtain
the equations of motion
\beq\label{te4a}
\begin{array}{lll}
\p_{\bar t_1}\p_{\bar t_1} x_i &=&\displaystyle{-\sum_{k\neq i}
\p_{\bar t_1} x_i \, \p_{\bar t_1}  x_k \Bigl (
\zeta (x_i-x_k+\eta )+\zeta (x_i-x_k-\eta )-2\zeta (x_i-x_k)\Bigr )}
\\ && \\
&=&\displaystyle{\sum_{k\neq i} \p_{\bar t_1} x_i \, \p_{\bar t_1}x_k
\frac{\wp '(x_i-x_k)}{\wp (\eta )-
\wp (x_i-x_k)}}
\end{array}
\eeq
which are the same as equations of motion (\ref{te4}) for $t_1$-dynamics.
They are Hamiltonian with the Hamiltonian
\beq\label{te5a}
\bar H=\sum_i e^{-p_i}\prod_{k\neq i}\frac{\sigma (x_i-x_k-\eta )}{\sigma (x_i-x_k)}.
\eeq
Taking into account (\ref{te6}) and the fact that
$$
\det_{ij} \Phi (x_i-x_j-\eta )=C(\eta , \lambda )\prod_{j<k}
\frac{\sigma^2(x_j-x_k)}{\sigma (x_j-x_k+\eta )\sigma (x_j-x_k-\eta )}
$$
with a constant $C(\eta , \lambda )$,
it is easy to see that
\beq\label{te8}
\bar H=-e^{-2c\eta^2}\sigma^{-2}(\eta )\sum_i \p_{\bar t_1}x_i=
\mbox{const}\, \mbox{tr}\, L^{-1}.
\eeq

\subsection{Self-dual form of the Ruijsenaars-Schneider equations of motion}

Substituting $\psi (x) = z^{x/\eta}e^{t_1 z}\hat \tau (x)/\tau (x)$ into the first
linear problem in (\ref{t2}), we get the equation
\beq\label{sdrs1}
\p_{t_1}\log \frac{\hat \tau (x)}{\tau (x+\eta )}=
z\, \frac{\hat \tau (x+\eta )\tau (x)}{\tau (x+\eta )\hat \tau (x)}-z.
\eeq
As before, we parametrize the function $\hat \tau$ by its zeros $y_i$, so that
$$
\frac{\hat \tau (x)}{\tau (x)}=e^{\alpha x+\beta t_1}\prod_{i}
\frac{\sigma (x-y_i)}{\sigma (x-x_i)}
$$
with some constants $\alpha , \beta$. Equation (\ref{sdrs1}) becomes
$$
\sum_i \Bigl (\dot x_i \zeta (x-x_i+\eta )-\dot y_i \zeta (x-y_i)\Bigr )
 =
e^{\mu}\prod_i \frac{\sigma (x-x_i)\sigma (x-y_i+\eta )}{\sigma (x-y_i)\sigma (x-x_i+\eta )}
+\mbox{const}
$$
with a constant $\mu$. Equating residues at $x=x_i-\eta$ and $x=y_i$, we get the 
following equations:
\beq\label{sdrs2}
\left \{
\begin{array}{l}
\displaystyle{\dot x_i=-\sigma (\eta )e^{\mu}\prod_{k\neq i}
\frac{\sigma (x_i-x_k-\eta )}{\sigma (x_i-x_k)}\prod_j 
\frac{\sigma (x_i-y_j)}{\sigma (x_i-y_j-\eta )}}
\\ \\
\displaystyle{\dot y_i=-\sigma (\eta )e^{\mu}\prod_{k\neq i}
\frac{\sigma (y_i-y_k+\eta )}{\sigma (y_i-y_k)}\prod_j 
\frac{\sigma (y_i-x_j)}{\sigma (y_i-x_j+\eta )}}.
\end{array}
\right.
\eeq
This is the self-dual form of the Ruijsenaars-Schneider equations of motion for
$x_i$ and $y_i$. 
For the proof that the latter follow from (\ref{sdrs2}) see \cite{ZZ18}.

The self-dual form of equations of motion is directly connected with the 
integrable time discretization of the Ruijsenaars-Schneider system \cite{NRK96,KWZ98}. 
In this interpretation, $\tau$ and $\hat \tau$ are tau-functions taken at two subsequent 
values $n$ and $n+1$ of the discrete time. Accordingly, we denote $x_i = x_i^n$,
$y_i=x_{i}^{n+1}$. It then follows from (\ref{sdrs2}) that the discrete time dynamics
is given by equations of motion
\beq\label{sdrs3}
\prod_{k=1}^{N}\frac{\sigma (x_i^n -x_k^{n-1})}{\sigma (x_i^n-x_k^{n-1}+\eta )}\,
\frac{\sigma (x_i^n -x_k^{n}+\eta )}{\sigma (x_i^n -x_k^{n}-\eta )}\,
\frac{\sigma (x_i^n -x_k^{n+1}-\eta )}{\sigma (x_i^n-x_k^{n+1})}=-1.
\eeq
Remarkably, equations (\ref{sdrs3}) coincide with the nested Bethe ansatz equations
for the generalized integrable magnet with elliptic $R$-matrix
associated with the root system $A_m$, with the 
discrete time $n$ taking values $0, 1, \ldots , m+1$. Equations (\ref{d3}) are reproduced
in the limit $\eta \to 0$.
For the Hamiltonian approach to the time discretization of integrable systems see
\cite{Suris}.

\section{Conclusion}

In this paper we have reviewed double-periodic (elliptic) solutions
to integrable non-linear partial differential equations (KP, BKP, 2DTL) and have presented
a detailed derivation of 
equations of motion for their poles. The dynamics of poles for KP and 2DTL equations
is given by the known integrable many-body systems (elliptic 
Calogero-Moser and Ruijsenaars-Schneider models
respectively) while for the BKP equation a new many-body system with three-body interaction
arises. It is an open question to prove integrability of this system and to find whether
or not it is Hamiltonian. We were able to find explicitly only a few non-trivial integrals
of motion for this system expressed through coordinates and velocities. 
We believe that the system is integrable since the equation
of the spectral curve depending on the spectral parameter provides a large supply 
of conserved quantities. 

We have also discussed the so-called self-dual form of 
equations of motion which is intimately connected with the integrable time discretization
of the many-body systems. The equations of motion in discrete time mysteriously coincide
with the nested Bethe ansatz equations for quantum integrable models with elliptic
$R$-matrix.
 
There is an open question even for the more familiar KP/Calogero-Moser case. 
It would be very desirable to extend Shiota's result \cite{Shiota94}
to the elliptic solutions. Namely, the problem is to establish the
correspondence between elliptic solutions to the KP equation and the Calogero-Moser
system with elliptic potential on the level of hierarchies, i.e., to prove that 
the evolution of poles
with respect to the higher times $t_k$ of the infinite KP hierarchy is governed by
higher Hamiltonians $H_k$ of the elliptic Calogero-Moser system (which are yet to be 
determined explicitly). In this paper we did this for $k=2$ and $k=3$.

\section{Appendices}

\subsection*{Matrix identities}

Here we prove certain useful identities for the off-diagonal matrices
$$A_{ij}=(1-\delta_{ij})\Phi (x_i-x_j), \quad
B_{ij}=(1-\delta_{ij})\Phi ' (x_i-x_j), \quad
C_{ij}=(1-\delta_{ij})\Phi '' (x_i-x_j)$$ and diagonal matrices
$$\displaystyle{D_{ij}=\delta_{ij}\sum_{k\neq i}\wp (x_i-x_k)}, \quad
\displaystyle{D'_{ij}=\delta_{ij}\sum_{k\neq i}\wp '(x_i-x_k)}, \quad
\displaystyle{D'''_{ij}=\delta_{ij}\sum_{k\neq i}\wp '''(x_i-x_k)}.$$

We begin with the identity
\beq\label{A4}
[A,B]+[A,D]=D'.
\eeq
To transform the commutators
$[A,B]+[A,D]$, we use the identity
\beq\label{A1}
\Phi (x )\Phi '(y)-\Phi (y)\Phi '(x)=\Phi (x+y)(\wp (x) -\wp (y))
\eeq
which, in turn, directly follows from the easily proved identity
\beq\label{A2}
\Phi (x, \lambda )\Phi (y, \lambda )=\Phi (x+y, \lambda )
\Bigl (\zeta (x)+\zeta (y)-\zeta (x+y+\lambda )+\zeta (\lambda )\Bigr ).
\eeq
With the help of (\ref{A1}) we get for $i\neq k$
$$
\Bigl ([A,B]+[A,D]\Bigr )_{ik}=\, \sum_{j\neq i,k}\Phi (x_i-x_j)\Phi '(x_j-x_k)-
\sum_{j\neq i,k}\Phi ' (x_i-x_j)\Phi (x_j-x_k)
$$
$$
+\, \Phi (x_i-x_k)\Bigl (\sum_{j\neq k}\wp (x_j-x_k)-\sum_{j\neq i}\wp (x_i-x_j)\Bigr )=0,
$$
so $[A,B]+[A,D]$ is a diagonal matrix. To find the diagonal matrix elements, we use the 
special case 
of (\ref{A1}) at $y=-x$ (obtained as the limit $y\to -x$):
\beq\label{A3}
\Phi (x)\Phi '(-x)-\Phi (-x)\Phi '(x)=\wp '(x).
\eeq
This leads to
$$
\Bigl ([A,B]+[A,D]\Bigr )_{ii}
$$
$$
=\,
\sum_{j\neq i}\Bigl (\Phi (x_i-x_j)\Phi ' (x_j-x_i)-\Phi ' (x_i-x_j)\Phi (x_j-x_i)\Bigr )
=\sum_{j\neq i}\wp '(x_i-x_j)=D'_{ii},
$$
so we finally obtain the matrix identity (\ref{A4}).

Combining the derivatives of (\ref{A1}) w.r.t. $x$ and $y$, we obtain the 
identities
\beq\label{A5}
\Phi (x)\Phi ''(y)-\Phi (y)\Phi ''(x)=2\Phi '(x+y)(\wp (x)-\wp (y))
+\Phi (x+y)(\wp '(x)-\wp '(y)),
\eeq
\beq\label{A6}
\Phi '(x)\Phi ''(y)-\Phi '(y)\Phi ''(x)=\Phi ''(x+y)(\wp (x)-\wp (y))
+\Phi '(x+y)(\wp '(x)-\wp '(y)).
\eeq
Their limits as $y\to -x$ are
\beq\label{A7}
\Phi (x)\Phi ''(-x)-\Phi (-x)\Phi ''(x)=0,
\eeq
\beq\label{A8}
\Phi '(x)\Phi ''(-x)-\Phi '(-x)\Phi ''(x)=-\frac{1}{6}\, \wp '''(x)-\wp (\lambda )\wp '(x).
\eeq
Using these formulas, it is not difficult to prove the following matrix identities:
\beq\label{A9}
[A,C]=2[D, B]+D'A+AD',
\eeq
\beq\label{A10}
[B,C]=[D,C]+D'B+BD'-\frac{1}{6}\, D''' -\wp (\lambda ) D'.
\eeq

Finally, we will prove the identity
\beq\label{B1}
Y+\frac{1}{2}[\dot X, C]-\frac{1}{2}\, \wp (\lambda )[\dot X, A]=-\tilde D ',
\eeq
where
$$
Y=B\dot X A -A \dot X B -[A, \tilde D], \quad
\tilde D_{ij}=\delta_{ij}\sum_{k\neq i} \dot x_k \wp (x_i-x_k), \quad
\tilde D'_{ij}=\delta_{ij}\sum_{k\neq i} \dot x_k \wp ' (x_i-x_k).
$$
We have 
$$
Y_{ii}=\sum_{j\neq i}\Bigl (\Phi '(x_{ij})\dot x_j \Phi (x_{ji})-
\Phi (x_{ij})\dot x_j \Phi '(x_{ji})\Bigr )=-\sum_{j\neq i}\dot x_j \wp '(x_{ij})=
-\tilde D'_{ii}
$$
due to (\ref{A3}). At $i\neq k$ we get, using (\ref{A1}):
$$
Y_{ik}=\! \sum_{j\neq i,k}\Bigl (\Phi '(x_{ij})\dot x_j \Phi (x_{jk}) \! -\!
\Phi (x_{ij})\dot x_j \Phi '(x_{jk})\Bigr )
+\Phi (x_{ik})\sum_{j\neq i}\dot x_j \wp (x_{ij}) \! -\!
\Phi (x_{ik})\sum_{j\neq k}\dot x_j \wp (x_{kj})
$$
$$
=\Phi (x_{ik})\sum_{j\neq i,k}\dot x_j \Bigl (\wp (x_{jk})-\wp (x_{ij})\Bigr )
+\Phi (x_{ik}) \Bigl (\sum_{j\neq i}\dot x_j \wp (x_{ij})-
\sum_{j\neq k}\dot x_j \wp (x_{kj})\Bigr )
$$
$$
=-(\dot x_i-\dot x_k)\Phi (x_{ik})\wp (x_{ik}).
$$
Now we have for $i\neq k$:
$$
Y_{ik}+\frac{1}{2}[\dot X, C]_{ik}-\frac{1}{2}\, \wp (\lambda )[\dot X, A]_{ik}=
\frac{1}{2}\, (\dot x_i-\dot x_k)\Bigl (\Phi ''(x_{ik})-\Phi (x_{ik})(2\wp (x_{ik})-
\wp (\lambda))\Bigr )=0
$$
due to (\ref{A15a}), see below in the appendix. Since $[\dot X, A]_{ii}=[\dot X, C]_{ii}=0$,
the matrix identity (\ref{B1}) is proved.

\subsection*{Proof of the identity used in section \ref{section:sdcm}}

Here we prove the identity
\beq\label{B2}
\begin{array}{l}
 -\displaystyle{\sum_{j\neq i}\Bigl (\sum_{k\neq i}\zeta (x_i-x_k)\! -\! 
\sum_k \zeta (x_i-y_k)\! -\! \sum_{k\neq j}\zeta (x_j-x_k)\! +\! \sum_k \zeta (x_j -y_k)\Bigr )
\wp (x_i-x_j)}
\\ \\
 +\displaystyle{\sum_{j}\Bigl (\sum_{k\neq i}\zeta (x_i-x_k)\! -\! 
\sum_k \zeta (x_i-y_k)\! +\! \sum_{k\neq j}\zeta (y_j-y_k)\! -\! \sum_k \zeta (y_j -x_k)\Bigr )
\wp (x_i-y_j)}
\\ \\
 -\, \displaystyle{\sum_{j\neq i}\wp '(x_i-x_j)}=0,
\end{array}
\eeq
where $x_1, \ldots , x_N$, $y_1, \ldots , y_N$ are arbitrary variables.

The first non-trivial case is $N=2$. Put $i=1$ and 
consider the left hand side as a function of $x_1$.
It is easy to see that it is an elliptic function of $x_1$. It may have poles at 
$x_1=x_2$, $x_1=y_1$, $x_1=y_2$. Setting $x_1=x_2 +\varepsilon$, we can directly check that
the left hand side is regular at $x_1=x_2$ and 
moreover it is $O(\varepsilon )$ as $\varepsilon \to 0$, so it vanishes at
$x_1=x_2$. In a similar way, 
one can check that the left hand side is regular at $x_1=y_1$, $x_1=y_2$.
It follows from these facts that it is identically equal to zero. 
The argument for $i=2$ is the same. 

Passing to the general case, 
let us denote the left hand side of (\ref{B2}) by 
$F^{(i)}_N=F^{(i)}_N(x_1, \ldots , x_N, y_1, \ldots , y_N)$ and consider it as a function
of $x_i$. It is easy to see that it is an elliptic function of $x_i$.
It may have poles at $x_i=x_{i_0}$ ($i_0=1,\ldots , N$, $i_0\neq i$) and
$x_i=y_{i_0}$ ($i_0=1,\ldots , N$). Setting $x_i=x_{i_0}+\varepsilon$,
$x_i=y_{i_0}+\varepsilon$, it can be 
checked that $F^{(i)}_N$ is regular, i.e., all singular contributions cancel 
and $F^{(i)}_N=O(1)$ as $\varepsilon \to 0$. Therefore, $F^{(i)}_N$ is a constant
independent of $x_i$. To find the constant, we expand $F^{(i)}_N$ around $x_{i_0}$
up to the constant term in $\varepsilon$:
\beq\label{B4}
F^{(i)}_N=F^{(i_0)}_{N-1}(x_1, \ldots , \hat x_i, \ldots , x_N, y_1, \ldots , 
\hat y_i , \ldots , y_N)
+G^{(i_0)}_{N-1} +O(\varepsilon ),
\eeq
where $\hat x_i$, $\hat y_i$ means that $x_i$, $y_i$ are omitted and
$$
G^{(i_0)}_{N-1}=G^{(i_0)}_{N-1}(x_1, \ldots , \hat x_i , \ldots , x_N, y_1, \ldots , y_N)
$$
is given by
$$
G^{(i_0)}_{N-1}=\frac{1}{2}\sum_k \wp '(x_{i_0}-y_k)-
\frac{1}{2}\sum_{k\neq i, i_0} \wp '(x_{i_0}-x_k)+\zeta (x_{i_0}-y_i)\wp (x_{i_0}-y_i)
$$
$$
-\sum_{j\neq i, i_0}\Bigl (\zeta (x_{i_0}-x_j)-\zeta (x_{i_0}-y_i)+\zeta (x_j-y_i)\Bigr )
\wp (x_{i_0}-x_j)
$$
$$
+\sum_{j\neq i}\Bigl (\zeta (x_{i_0}-y_j)-\zeta (x_{i_0}-y_i)+\zeta (y_j-y_i)\Bigr )
\wp (x_{i_0}-y_j)
$$
$$
+\Bigl (\sum_{k\neq i, i_0}\zeta (x_{i_0}-x_k)-\sum_{k\neq i}\zeta (x_{i_0}-y_k)
+\sum_{k\neq i} \zeta (y_i-y_k)-\sum_{k\neq i, i_0}\zeta (y_i-x_k)\Bigr )\wp (x_{i_0}-y_i).
$$
In the second and the third lines we use the identity
\beq\label{B3}
\zeta (x)-\zeta (y)-\zeta (x-y)=-\frac{1}{2}\, \frac{\wp '(x)+\wp '(y)}{\wp (x)-\wp (y)}
\eeq
to get that the sum of the second and the third lines is
$$
\sum_{j\neq i, i_0}
\left (\frac{1}{2}\, \wp '(x_{i_0}-x_j)-\frac{1}{2}\, \wp '(x_{i_0}-y_j)\right.
$$
$$
\left.
+\Bigl (\zeta (x_{i_0}-y_j)+\zeta (y_j-y_i)-\zeta (x_{i_0}-x_j)-\zeta (x_j-y_i)\Bigr )
\wp (x_{i_0}-y_i)\phantom{\frac{1}{2}}\!\!\! \right )
$$
$$
+\Bigl (\zeta (x_{i_0}-y_{i_0})-\zeta (x_{i_0}-y_i)+\zeta (y_{i_0}-y_i)\Bigr )
\wp (x_{i_0}-y_{i_0}).
$$
Substituting this back into the expression for $G^{(i_0)}_{N-1}$, we get after some
cancellations:
$$
G^{(i_0)}_{N-1}=\frac{1}{2}\, \wp '(x_{i_0}-y_i)+\frac{1}{2}\, \wp '(x_{i_0}-y_{i_0})
$$
$$
+\Bigl (\zeta (x_{i_0}-y_{i_0})-\zeta (x_{i_0}-y_i)+\zeta (y_{i_0}-y_i)\Bigr )
\Bigl (\wp (x_{i_0}-y_{i_0})-\wp (x_{i_0}-y_{i})\Bigr ).
$$
Using again the identity (\ref{B3}), we see that $G^{(i_0)}_{N-1}=0$. 

Now we are going to use the inductive argument: suppose that $F_{N-1}^{(i)}=0$
(this is true for $N=3$), then we see from (\ref{B4}) that 
$F_{N}^{(i)}=O(\varepsilon )$ as $\varepsilon \to 0$ and, therefore, 
$F_{N}^{(i)}=0$. 

\subsection*{Proof of equation (\ref{a3})}

Here we prove the identity (\ref{a3}).
Using the explicit form of the matrices $L$, $M$ (\ref{a1a}), (\ref{a1b}), we write
$$
\begin{array}{lll}
\dot L+[L,M]&=&36z^2 \Bigl ([A,B]+[A,D]\Bigr )
\\ && \\
&&-\, 6z \Bigl (\dot A -[\dot X, B]\Bigr )+\, 36 z\Bigl ([A,C]-[A, D']+2[B, D]\Bigr )
\\ &&\\
&&-\, 6 \Bigl (\dot B -[\dot X, C]\Bigr )-\ddot X +6\dot D
\\ &&\\
&&+\, 36 \Bigl ([B,C]-[B, D']+[C, D]\Bigr ).
\end{array}
$$
First of all we notice that $\dot A_{ik}=(\dot x_i-\dot x_k)\Phi '(x_i-x_k)$,
$\dot B_{ik}=(\dot x_i-\dot x_k)\Phi ''(x_i-x_k)$, and, therefore, 
we have $\dot A =[\dot X, B]$, $\dot B =[\dot X, C]$. Next, we have
$[A,B]+[A,D]=D'$ (see (\ref{A4})) and equations (\ref{A9}), (\ref{A10})
are used to transform $\dot L+[L,M]$ to the form (\ref{a3}).

\subsection*{Some useful identities}

Apart from already mentioned identities for the $\Phi$-function for the calculations 
in Sections 2 and 3 we need 
the following ones:
\beq\label{A11}
\Phi (x)\Phi (-x)=\wp (\lambda )-\wp (x),
\eeq
\beq\label{A12}
\Phi '(x)\Phi (-x)+\Phi '(-x)\Phi (x)=\wp '(\lambda ),
\eeq
\beq\label{A13}
\Phi '(x)\Phi ' (-x)=\wp ^2(x)+\wp (\lambda )\wp (x)+\wp ^2(\lambda )-
\frac{1}{4}\, g_2,
\eeq
\beq\label{A14}
\Phi (x)\Phi '' (-x)=\wp ^2(\lambda )+\wp (\lambda )\wp (x)-2\wp ^2(x ),
\eeq
\beq\label{A15}
\Phi '(x)\Phi '' (-x)=\Bigl (\wp '(\lambda )-\wp '(x)\Bigr )
\Bigl (\wp (x)+\frac{1}{2}\, \wp (\lambda )\Bigr ),
\eeq
\beq\label{A15a}
\Phi ''(x)=\Phi (x)(2\wp (x)-
\wp (\lambda)).
\eeq
They eventually follow from the basic identity (\ref{A2}). We also need some 
well known identities for the Weierstrass functions:
\beq\label{A16}
2\zeta (\lambda )-\zeta (\lambda +x)-\zeta (\lambda -x)=
\frac{\wp '(\lambda )}{\wp (x)-\wp (\lambda )},
\eeq
\beq\label{A16a}
\wp '^2(x)=4\wp ^3(x)-g_2\wp (x)-g_3,
\eeq
\beq\label{A17}
\wp (x+\lambda )-\wp (x-\lambda )=-\, \frac{\wp '(\lambda )\wp '(x)}{(\wp (x)-
\wp (\lambda ))^2},
\eeq
\beq\label{A18}
\wp (x+\lambda )+\wp (x-\lambda )=\frac{1}{2}\,
\frac{\wp '^2(x)+\wp '^2(\lambda )}{(\wp (x)-
\wp (\lambda ))^2}-2\Bigl (\wp (x)+\wp (\lambda )\Bigr ).
\eeq

\section*{Acknowledgments}
\addcontentsline{toc}{section}{\hspace{6mm}Acknowledgments}

The author thanks I. Krichever, S. Natanzon, D. Rudneva and P. Wiegmann for discussions.
The work was performed at the Steklov Mathematical Institute of Russian 
Academy of Sciences, Moscow. 
This work is supported by the Russian Science Foundation under grant 19-11-00062.

\end{document}